\documentclass[11pt]{article}

\usepackage[margin=1in]{geometry}
\usepackage{mathtools}
\usepackage{amssymb}
\usepackage{physics}
\usepackage{chemformula}
\usepackage{algorithm}
\usepackage{algpseudocode}
\usepackage{booktabs}
\usepackage{pgfplots}
\pgfplotsset{compat=1.18}
\usepackage{siunitx}
\usepackage[hidelinks]{hyperref}
\usepackage{makecell}
\usepgfplotslibrary{groupplots}
\newcommand{\norb}{N}
\newcommand{\morb}{M}
\newcommand{\nelec}{n}
\newcommand{\NFCI}{N_\text{FCI}}
\newcommand{\occ}{\text{occ}}
\newcommand{\Iset}{\mathcal{I}_{H}}
\newcommand{\oneD}{{}^1D}
\newcommand{\twoD}{{}^2D}

\newcommand{\diff}{\mathrm{d}}
\newcommand{\T}{\top}
\newcommand{\E}{\mathcal{E}}
\newcommand{\I}{\mathcal{I}}
\newcommand{\bigO}{\mathcal{O}}
\newcommand{\argmax}{\mathop{\arg\max}}
\newcommand{\argmin}{\mathop{\arg\min}}
\newcommand\orth[1]{\,\mathrm{orth}\bigl(#1\bigr)}

\renewcommand{\op}[1]{\hat{\boldsymbol{#1}}}
\renewcommand{\vec}[1]{\boldsymbol{#1}}

\newcommand{\ham}{\op{H}}
\newcommand{\vvc}{\vec{c}}
\newcommand{\vvb}{\vec{b}}
\newcommand{\vve}{\vec{e}}
\newcommand{\currvc}{\vvc^{(\ell)}}
\newcommand{\currvb}{\vvb^{(\ell)}}
\newcommand{\nextvc}{\vvc^{(\ell+1)}}
\newcommand{\nextvb}{\vvb^{(\ell+1)}}
\newcommand{\ichosen}{i^{(\ell+1)}}
\newcommand{\Ichosen}{\I^{(\ell+1)}}
\newcommand{\achosen}{\eta^{(\ell+1)}}
\newcommand{\gachosen}{\gamma^{(\ell+1)}}
\newcommand{\vachosen}{\vec{\eta}^{(\ell+1)}}
\newcommand{\ccnorm}{\mu}   
\newcommand{\ccquad}{\alpha}   
\newcommand{\currC}{C^{(\ell)}}
\newcommand{\currB}{B^{(\ell)}}
\newcommand{\currG}{G^{(\ell)}}
\newcommand{\nextC}{C^{(\ell+1)}}
\newcommand{\nextB}{B^{(\ell+1)}}
\newcommand{\CCnorm}{M}   
\newcommand{\CCquad}{A}   
\newcommand{\NH}{N_{H}}

\usepackage{authblk}

\title{CDFCI: High-Performance Parallel Software for Quantum Many-Body Eigenvalue Problems}

\author[1]{Yuejia Zhang}
\author[2]{Zhe Wang}
\author[2,3,4]{Jianfeng Lu}
\author[1,5,6]{Yingzhou Li}

\affil[1]{School of Mathematical Sciences, Fudan University, Shanghai, China}
\affil[2]{Department of Mathematics, Duke University, Durham, NC, USA}
\affil[3]{Department of Physics, Duke University, Durham, NC, USA}
\affil[4]{Department of Chemistry, Duke University, Durham, NC, USA}
\affil[5]{Shanghai Key Laboratory for Contemporary Applied Mathematics, Shanghai, China}
\affil[6]{Key Laboratory of Computational Physical Sciences, Ministry of Education, China}

\date{}

\begin{document}

\maketitle

\begin{abstract}
CDFCI is a shared-memory parallel numerical program for computing low-lying
eigenpairs of large-scale, non-relativistic fermionic Hamiltonians.
The software is designed to handle a broad class of quantum many-body models,
including both \textit{ab initio} electronic structure Hamiltonians and 
lattice model Hamiltonians arising in condensed matter physics.
CDFCI combines an efficient coordinate-descent-based selected configuration
interaction algorithm with dedicated parallelization strategies, achieving
high performance on modern multi-core architectures.
The package implements several variants of the coordinate descent framework,
including CDFCI, mCDFCI, xCDFCI, and OptOrbFCI, supporting both ground- and
excited-state calculations, as well as the compression and optimization of
single-particle orbitals.
Benchmark results on representative quantum chemistry and condensed matter
test cases demonstrate that CDFCI attains state-of-the-art accuracy with
competitive performance compared to established selected configuration
interaction (such as CIPSI or SHCI) and DMRG implementations.
The software is open-source, extensively documented, and provides a Python
interface for seamless integration with PySCF and other many-body simulation
workflows.
\end{abstract}


\section{Introduction}
CDFCI is a software package designed to provide approximate
numerical solutions to the fermionic, time-independent, non-relativistic
many-body Schr\"{o}dinger equation.
This class of problems underpins computational physics and chemistry,
enabling the study of reactivity, spectroscopy, and functional properties
of molecular and condensed systems.
For most practically relevant problems,
the central task is to determine the low-lying eigenvalues
and eigenfunctions of the Hamiltonian operator
that describes the total energy of the system.
It is, however, intrinsically difficult to solve the many-body problem,
because neither analytic solutions exist
nor are numerical methods tractable,
due to the exponential growth of problem size with the number of particles.
There are system-level simplifications, including the
Hartree--Fock (HF) method and Density Functional Theory (DFT).
The former uses a mean-field approximation of particle interactions
and reduces the problem to an effective single-particle one,
while the latter directly works with the electron density instead of
the wavefunction.
In both cases, correlation effects are not treated explicitly,
which could lead to inaccurate results for strongly correlated systems.
Thus, methods that preserve the many-body wavefunction, also called
wavefunction-based methods,
remain crucial for treating strong correlation and are the focus of this paper.

One typical wavefunction-based method,
known as full configuration interaction (FCI) in quantum
chemistry,
refers to the process of expanding the many-body wavefunction
into a linear combination of many-body basis functions.
These many-body basis functions are constructed as antisymmetric tensor products of one-body basis functions,
with the aim of enforcing the Pauli exclusion principle.
They describe the occupation state of particles and are referred to as configurations throughout this paper.
With a given truncated one-body basis set,
the FCI discretization procedure
reduces the problem to a linear eigenvalue problem
whose dimension equals that of the
Hilbert space spanned by all possible configurations.
Solving such large, sparse and symmetric eigenvalue problems is standard practice
in numerical linear algebra, and iterative methods
such as the Davidson algorithm~\cite{Davidson1975}
have been widely employed
to compute the ground state and a few low-lying excitations.
Nevertheless, the exponential growth of the Hilbert space
soon renders these established techniques
impractical for realistic systems.
Recent FCI calculations have reached determinant spaces at the trillion and even quadrillion scale~\cite{trillion2024, quadrillion2025}.
A variety of specialized approaches have therefore been developed,
which are commonly divided into categories including
selected configuration interaction with perturbation theory,
quantum Monte Carlo, density matrix renormalization group and
others.

Perturbation theory (PT) inspires the splitting of the Hamiltonian
into a solvable reference term and the rest as a perturbation.
Correlation effects can be captured at modest additional cost
via low-order corrections (e.g., MP2/MPn), although accuracy
depends sensitively on the choice of the reference Hamiltonian
and the weakness of the perturbation~\cite{Moller1934}.
A classic method of this type in electronic structure calculations
is configuration interaction singles-doubles (CISD),
where the reference space includes the Hartree--Fock state
and all single- and double-excitation states built from it~\cite{Pople1977}.
In the meantime, selected configuration interaction (selected CI) methods emerged,
constructing the solution space by adding determinants with the
largest contributions iteratively, typically guided by a
second-order perturbative estimate (PT2).
From the seminal CIPSI approach~\cite{Huron1973},
the selected CI framework has evolved into a broad family of methods,
including more sophisticated schemes such as
adaptive sampling CI~\cite{tubmanASCI1, minASCI3},
heat-bath CI~\cite{HolmesHCI, ugandiHCI3},
semistochastic heat-bath CI~\cite{SharmaSHCI, wangSHCI3}
that incorporate semistochastic PT2 corrections,
iterative CI with selection~\cite{zhangiCI},
fast randomized iteration method for FCI~\cite{Greene2020},
reinforcement learning CI~\cite{Goings2021},
together with large-scale parallel and GPU-accelerated
software implementations~\cite{williamsyoungASCI2,dangHCI2,liSHCI2,garnironQuantumPackage2019}.

Quantum Monte Carlo (QMC) methods offer another
approach for sampling the wavefunction, but suffer from the notorious
fermion sign problem~\cite{Loh1990}.
The full configuration interaction quantum Monte Carlo (FCIQMC) method
introduced by Booth et al.~\cite{Booth2009}
mitigates this issue by employing a population dynamics algorithm
to evolve a set of walkers in the configuration space~\cite{Spencer2012},
and further enhancements, such as
the initiator method~\cite{Cleland2010FCIQMC}
and semi-stochastic projector techniques~\cite{Petruzielo2012, Blunt2015},
significantly improve its efficiency and robustness across
a wide range of molecular and condensed-phase
systems~\cite{Booth2013Nature,Guther2020NECI}.
Density matrix renormalization group (DMRG), on the other hand,
reformulates the problem in terms of matrix product states, enabling efficient
treatment of one-dimensional or quasi-one-dimensional systems,
and has been successfully adapted to \textit{ab initio} quantum chemistry
via suitable orbital orderings and active
spaces~\cite{White1993,WhiteMartin1999,schollwockDMRG,sharmaDMRG}.
Efficient implementations of the DMRG method have achieved
cutting-edge performance results and nearly ideal parallel scaling
on high-performance computing platforms~\cite{zhaiLowCommunicationHigh2021,
levyDistributedMemoryDMRGSparse2020,menczerParallelImplementationDensity2024}.
Finally, many other approximation theories and methods have been developed,
such as the FCC reduction method combining full coupled cluster theory (FCC)
with FCI solutions~\cite{xuNearExactSolutionsMolecular2020},
and the many-body expansion FCI approaches (MBE-FCI) based on low-order
truncations of many-body expansions~\cite{eriksenManyBodyExpandedFull2018,
eriksenManyBodyExpandedFull2019}.

This paper introduces the software package CDFCI
(short for Coordinate Descent Full Configuration Interaction),
which implements a series of methods based on the core idea of
transforming the original eigenvalue problem into
an equivalent unconstrained optimization problem,
and employing a customized coordinate gradient descent framework.
Each coordinate corresponds to a configuration,
or, more fundamentally, a basis function in Hilbert space.
Our method is an iterative approach
that progressively expands the variational space ---
selectively including coordinates with high contributions
based on the magnitude of their gradient entries.
Unlike conventional subspace iterative eigensolvers
such as the Lanczos and Davidson methods, which can also be implemented
in a matrix-free manner, CDFCI avoids global matrix-vector products
and instead maintains and updates sparse representations of the active
many-body wavefunctions through coordinate-wise operations.
The action of the Hamiltonian is updated incrementally,
requiring access only to matrix elements associated with the selected
coordinates.
This selective access is particularly advantageous when the determinant
space is enormous but only a small fraction of configurations contribute
significantly to the target wavefunctions.

CDFCI also differs from conventional selected CI methods by
directly selecting and updating wavefunction coefficients according to
the magnitude of their gradient entries, rather than repeatedly solving
the eigenvalue problem in an expanding variational subspace.
This optimization perspective provides a theoretical foundation for
the convergence analysis of coordinate-descent FCI methods
under appropriate assumptions~\cite{li2026convergenceerroranalysiscoordinate}.
At the end of all iterations, the lower spectrum of the Hamiltonian
is estimated in a threshold-controlled subspace,
which can be further corrected when combined with perturbation theory.
The CDFCI software package is publicly available at
\url{https://github.com/CDFCI/CDFCI/}, and all numerical experiments
reported in this paper can be reproduced using the released code and
input examples.

The original CDFCI method~\cite{wangCoordinateDescentFull2019},
proposed by Wang, Li, and Lu in 2019, was used to solve for the ground state
in electronic structure calculations.
With theoretical convergence
guarantees~\cite{liCoordinateWiseDescentMethods2019}, the algorithm
follows a coordinate descent framework and performs an exact line search.
Later, Wang et al.\ proposed the
xCDFCI method~\cite{wangCoordinateDescentFull2023},
which modified the objective function to also
target low-lying excited states,
while the similar coordinate descent routine was employed.
In a series of works surrounding CDFCI,
mCDFCI~\cite{zhangParallelMulticoordinateDescent2025}
extended the single coordinate descent iteration to multiple coordinates.
The improvement of the parallel efficiency owes to
increasing workloads per step, which makes shared-memory parallelization
more flexible and balanced on multi-core machines.
Another related work is OptOrbFCI~\cite{liOptimalOrbitalSelection2020},
proposed by Li et al., which allows one-electron orbitals to be further rotated
or compressed.
The compression of orbitals in the outer loop
is combined with the CDFCI iteration in the inner loop,
thereby approximating optimal solutions in the variational space within
limited memory budget, similar to ideas of
complete active space self-consistent field (CASSCF) methods~\cite{Roos1987}.

The rest of the paper is organized as follows.
In Section~\ref{sec:problem}, we introduce the
formalism of the time-independent Schr\"{o}dinger equation
in a discretized basis set for fermions
and the reformulation of the eigenvalue problem.
The coordinate descent framework
and all numerical methods incorporated in the current
CDFCI package can be found in
Section~\ref{sec:methodology}.
Following this, Section~\ref{sec:software} is dedicated
to the practical side, including implementations and software designs,
with an illustrative example
to demonstrate the basic usage. Finally, Section~\ref{sec:experiments}
presents up-to-date numerical results conducted on
multi-core machines,
showcasing the accuracy and performance of the software, and
Section~\ref{sec:conclusion} concludes this paper.

\section{Problem Formulation}
\label{sec:problem}
This section presents the theoretical and algorithmic foundations
of our approach. We begin by defining
the discretized variational space that we consider
for the time-independent Schr\"{o}dinger equation
in Section~\ref{sec:basis-set}, and
reviewing the formulation of fermionic
Hamiltonians in the second quantization formalism in
Section~\ref{sec:hamiltonian-representations},
along with typical examples in electronic structure and lattice models.
This provides the necessary physical and mathematical background,
before we finally present the problem setup in
Section~\ref{sec:problem-formulation}, including the core idea of
reformulating the eigenvalue problem to an optimization problem.

\subsection{FCI Variational Space}
\label{sec:basis-set}
In quantum mechanics, the Hamiltonian operator \(\ham\) represents the total
energy of a quantum system, comprising both kinetic and potential
contributions. Its spectrum determines many fundamental
properties of the system, as captured by the time-independent Schr\"{o}dinger
equation,
\begin{equation}
  \label{eq:se-op}
  \ham \Psi = E \Psi.
\end{equation}
Here, \(\Psi\) denotes the many-body wavefunction
representing an eigenstate of the system with eigenvalue \(E\).
Let \(\nelec\) be the number of fermions.
According to quantum mechanics,
\(\Psi(\vec x_1,\dots,\vec x_\nelec):(\mathbb{R}^3\times\{\uparrow,\downarrow\})^\nelec\to\mathbb{C}\)
is a complex-valued function of the spatial and spin coordinates of the \(\nelec\) fermions and \(\Psi\)
belongs to a complex Hilbert space \(\mathcal{H}\).
We restrict attention to 
\(\bigwedge^{\nelec}
  L^2(\mathbb{R}^3\times\{\uparrow,\downarrow\},\mathbb{C}),\)
the antisymmetric subspace of \(\mathcal{H}\),
which both simplifies the computations and enforces fermionic antisymmetry.

Let \(\{\phi_i\}_{i\in \mathbb{N}}\) be a countable orthonormal basis
of \(L^2(\mathbb{R}^3\times\{\uparrow,\downarrow\},\mathbb{C})\). In practical computations, we restrict to a finite
subset \(\{\phi_1, \dots, \phi_\norb\}\).
These single-particle basis functions
are referred to as spin--orbitals in electronic structure theory and as
localized site states in lattice models.
Exploiting the isomorphism \(L^2(\mathbb{R}^3\times\{\uparrow,\downarrow\},\mathbb{C}) \cong L^2(\mathbb{R}^3,\mathbb{C})\otimes\mathbb{C}^2\), one distinguishes
spatial orbitals, which are functions in \(L^2(\mathbb{R}^3, \mathbb{C})\)
that do not incorporate the spin degree of freedom.

The finite-dimensional antisymmetric many-fermion space induced by the
chosen spin--orbitals is
\begin{equation}
  \label{eq:fci-variational-space}
  \mathcal{H}_{\mathrm{FCI}}
  \coloneqq
  \bigwedge^{\nelec}\operatorname{span}\{\phi_1,\dotsc,\phi_\norb\}
  =
  \operatorname{span}
  \left\{
    \phi_{i_1}\wedge\dotsb\wedge\phi_{i_\nelec}
    : 1\leq i_1<\dotsb<i_\nelec\leq\norb
  \right\}.
\end{equation}
The space \(\mathcal{H}_{\mathrm{FCI}}\) is
conventionally termed the full configuration interaction
(FCI) space in electronic structure calculations,
with dimension \(\binom{\norb}{\nelec} \eqqcolon \NFCI\).
It is alternatively called the variational space
because the associated problem is formulated
within a variational framework.
A canonical orthonormal basis of \(\mathcal{H}_{\mathrm{FCI}}\)
is furnished by the Slater determinants
\(\{\Phi_{i_1, \dotsb, i_\nelec} = \phi_{i_1} \wedge \dotsb \wedge \phi_{i_{\nelec}} :
1 \le i_1 < \dotsb < i_\nelec \le \norb\}\).

\subsection{Hamiltonian Representations}
\label{sec:hamiltonian-representations}
Working in Fock space, it is natural to adopt the second quantization
formalism, in which the Hamiltonian admits a compact algebraic form:
a general fermionic Hamiltonian
can be written as a linear combination of ladder operators that obey
anti-commutation relations. We consider the Hamiltonian
of the following form:
\begin{equation}
  \label{eq:general-ham}
  \ham = \sum_{p,q}^\norb h_{pq} \hat{a}_{p}^\dagger \hat{a}_{q}
  + \frac{1}{2} \sum_{p,q,r,s}^\norb v_{pqrs}
  \hat{a}_{p}^\dagger \hat{a}_{q}^\dagger \hat{a}_{s} \hat{a}_{r},
\end{equation}
where \(\hat{a}_{p}^\dagger\) and \(\hat{a}_{q}\) are
fermionic creation and annihilation operators that
satisfy \(\{\hat{a}_{p}, \hat{a}_{q}^\dagger\} = \delta_{pq}\),
in which \(\{\cdot,\cdot\}\) denotes the anticommutator, i.e.,
\(\{\hat{A},\hat{B}\}=\hat{A}\hat{B}+\hat{B}\hat{A}\).
Although the Hamiltonian may,
in principle,
contain higher-body interactions,
in this work
we focus exclusively on one-
and two-body terms, which is sufficient
to describe non-relativistic fermionic Hamiltonians.

Two widely studied examples of this general form~\eqref{eq:general-ham}
illustrate its versatility.
The first is the \textit{ab initio} electronic Hamiltonian
under the Born--Oppenheimer approximation, where
\(h_{pq}\) and \(v_{pqrs}\) are referred to as
one- and two-body integrals respectively,
\begin{align}
    h_{pq} &= \int \diff \vec{r}_1 \, 
    \phi^\star_p(\vec{r}_1) \op{h}(\vec{r}_1) \phi_q(\vec{r}_1), 
    \quad p, q = 1, \dotsc, \norb, \label{eq:1int} \\
    v_{pqrs} &= \int \diff \vec{r}_1 \diff \vec{r}_2 \,
    \phi^\star_p(\vec{r}_1) \phi^\star_q(\vec{r}_2)
    \op{v}(\vec{r}_1, \vec{r}_2) \phi_r(\vec{r}_1) \phi_s(\vec{r}_2), 
    \quad p, q, r, s = 1, \dotsc, \norb. \label{eq:2int}
\end{align}
The one-body operator \(\op{h}(\vec{r}_1)\) and
the two-body operator \(\op{v}(\vec{r}_1, \vec{r}_2)\)
take in the coordinates of one electron and two electrons respectively:
\begin{align}
    \op{h}(\vec{r}_1) &= -\frac{1}{2}\nabla_{\vec{r}_1}^2
    - \sum_{A=1}^{N_{\text{nuc}}}
    \frac{Z_A}{\|\vec{r}_1-\vec{r}_A\|}, \\
    \op{v}(\vec{r}_1, \vec{r}_2) &= \frac{1}{\|\vec{r}_1 - \vec{r}_2\|},
\end{align}
with \(N_{\text{nuc}}\) as the number of nuclei in the system, and
\(Z_A\) as the charge number of the nucleus \(A\).

The second example is the Hubbard Hamiltonian,
\begin{equation}
  \ham = -t \sum_{\langle i,j \rangle, \sigma}
  (\hat{a}_{i\sigma}^\dagger \hat{a}_{j\sigma} + \text{h.c.} )
  + U \sum_i \hat{n}_{i\uparrow} \hat{n}_{i\downarrow},
\end{equation}
where \(\langle i,j \rangle\) denotes
neighboring sites \(i\) and \(j\) on the lattice,
and \(\sigma \in \{\uparrow, \downarrow\}\) is the spin index.
The first term describes
the kinetic delocalization of electrons between neighboring sites,
and the second term represents onsite repulsion which
penalizes double occupancy.
Parameters \(t\) and \(U\) control the strength of
hopping and interaction respectively, and their ratio
\(U/t\) determines whether the system behaves more
like a metal or an insulator.
The particle number operator \(\hat{n}_{i\sigma}\) is defined
as \(\hat{a}_{i\sigma}^\dagger \hat{a}_{i\sigma}\).
Consequently, the onsite repulsion term can be
identified as the two-body interactions
in the general Hamiltonian form~\eqref{eq:general-ham}.

\subsection{Optimization Problem for Eigenvalues}
\label{sec:problem-formulation}
As described in the previous section,
the target wavefunction \(\Psi\) will be sought
in \(\mathcal{H}_\text{FCI}\), spanned by the
Slater determinants \(\{\Phi_I\}_{I=1}^{\NFCI} =
\{\Phi_{i_1, \dotsb, i_\nelec} :
1 \le i_1 < \dotsb < i_\nelec \le \norb\}\) of finite one-particle
basis functions \(\{\phi_1, \dots, \phi_\norb\}\).
In this work, we restrict the
one-particle basis functions to be
real-valued, as the systems under
consideration generally admit
real-valued eigenfunctions.
Expressing the wavefunction \(\Psi\) as
\(\Psi = \sum_{I=1}^{\NFCI} c_I \Phi_I\) leads to
the matrix representation of the time-independent
Schr\"{o}dinger equation~\eqref{eq:se-op} as
\begin{equation}
  \label{eq:se-mat}
  H \vvc = E \vvc
\end{equation}
where \(H\) is real symmetric and of dimension \(\NFCI \times \NFCI\),
with each entry
\begin{equation}
  \label{eq:entry}
  H_{I,J} =
  \begin{cases}
    \sum_{p\in \occ(\Phi_I)} h_{pp} + \frac{1}{2} \sum_{p, q\in \occ(\Phi_I)}
      (v_{pqpq} - v_{pqqp}), & \Phi_I = \Phi_J, \\
      (\pm)
    \left[h_{rp} + \sum_{k\in \occ(\Phi_J)} (v_{rkpk} - v_{rkkp})\right], &
      \Phi_I = \hat{a}_{r}^\dagger \hat{a}_{p} \Phi_J,\\
      (\pm)
    (v_{rspq} - v_{rsqp}), & \Phi_I = \hat{a}_{r}^\dagger \hat{a}_{s}^\dagger
      \hat{a}_{p} \hat{a}_{q} \Phi_J, \\
    0, & \text{otherwise,}
  \end{cases}
\end{equation}
under the assumption that \(\ham\) follows the general form
\eqref{eq:general-ham}. Here,
\(\occ(\Phi_I) = \{i_1, \dotsc, i_n\}\) if
\(\Phi_I = \phi_{i_1} \wedge \dotsb \wedge \phi_{i_n}\).
The sign \((\pm)\)
denotes the phase factor arising from fermionic antisymmetry,
determined by the number of orbital permutations required to excite \(\Phi_J\) to \(\Phi_I\).
Each entry of the vector \(\vvc\) corresponds to the coefficient \(c_I\)
for basis function \(\Phi_I\).

Consider the following
unconstrained nonconvex optimization problem,
which aims to find the best rank-one symmetric negative semidefinite
approximation of the Hamiltonian matrix \(H\),
\begin{equation}
\label{eq:optimization}
    \min_{\vvc \in \mathbb{R}^{\NFCI}} \,
    f(\vvc; H) \coloneqq
    \|H+\vvc\vvc^\T\|^2_\text{F}.
\end{equation}
As analysed in~\cite{liCoordinateWiseDescentMethods2019},
\(\pm \sqrt{-E_0} \vec{v}_0 \)
are the only two local minimizers of this problem,
where \(\vec{v}_0\) is the normalized eigenvector corresponding to the
smallest non-degenerate eigenvalue \(E_0 < 0\).
Thus, solving the optimization problem~\eqref{eq:optimization}
reveals the ground-state energy \(E_0\) and the ground-state
wavefunction coefficient vector \(\vec{v}_0\).

When targeting not only the ground state \(\Psi_0\),
but also several
low-lying excited states \(\Psi_1, \dotsc, \Psi_K\),
the respective states are expanded in terms of basis functions
\begin{equation}
  \Psi_k = \sum_{I=1}^{\NFCI} C_{I,k} \Phi_I, \quad k = 0, 1, \dotsc, K
\end{equation}
and coefficients \(C_{I,k}\) form an orthonormal matrix \(C\) of size
\(\NFCI \times (K+1)\). The optimization problem can subsequently be
generalized to a multi-state formulation
\begin{equation}
\label{eq:optimization-K}
    \min_{C \in \mathbb{R}^{\NFCI \times (K+1)}} \,
    f(C;H) \coloneqq
    \|H+CC^\T\|^2_\text{F},
\end{equation}
thereby revealing several low-lying eigenpairs simultaneously.

Assume that the smallest eigenvalues of \(H\)
satisfy \(E_0 \le E_1 \le \dotsb \le E_K < 0\) and \(E_K < E_{(K+1)}\).
It was reported in~\cite{gaoTriangularizedOrthogonalizationFreeMethod2022}
that all local minima are global minima in this problem,
and they admit the form \(V\sqrt{\Lambda} Q\) where the matrix \(\Lambda\) is a
diagonal matrix with diagonal entries of \(-E_0, \dotsc, -E_K\),
the matrix \(V\) satisfies \(V^\T HV = \Lambda\), and the matrix \(Q\) is an arbitrary orthogonal matrix of size
\((K+1) \times (K+1)\).

\section{Coordinate Descent Methods}
\label{sec:methodology}
This section elaborates on the application of
coordinate gradient descent methods to
address problem~\eqref{eq:optimization}
and~\eqref{eq:optimization-K},
in order to reveal the ground state
as well as a few excited states of the system.
As a first-order optimization algorithm, the coordinate gradient
descent method is particularly suitable for high-dimensional
problems~\cite{Wright2015}. More importantly,
it naturally supports a matrix-free implementation in our problem,
since it only requires coordinate-wise access to matrix-vector
products and thus avoids explicitly forming or storing
the Hamiltonian matrix \(H\), whose dimension is prohibitively large.
Coordinate gradient descent proceeds as follows.
At each iteration, one coordinate
of the optimization variable is updated.
The selection strategy may be cyclic, stochastic or based on
the Gauss--Southwell rule, which identifies the coordinate
with the largest absolute gradient component. The step size is
subsequently determined either as a constant,
as part of a diminishing sequence
or via a line search procedure.
After the selected coordinate is updated,
this procedure is repeated until convergence.

In the following
sections, we describe all the
coordinate-descent-based solvers that are
currently implemented in the package, including
the ground state solver CDFCI~\cite{wangCoordinateDescentFull2019}
in Section~\ref{sec:cdfci},
the multicoordinate ground state solver
mCDFCI~\cite{zhangParallelMulticoordinateDescent2025}
in Section~\ref{sec:mcdfci},
and the excited-state solver
xCDFCI~\cite{wangCoordinateDescentFull2023}
in Section~\ref{sec:xcdfci}.
To address the constraints of limited memory,
an additional solver OptOrbFCI~\cite{liOptimalOrbitalSelection2020}
has been implemented for the alternating optimization of one-particle basis functions and wavefunction coefficients,
and the detailed description is provided in Section~\ref{sec:optorbfci}.

\subsection{Coordinate-Descent Methods for Ground State}
\label{sec:cdfci}
At its core, the CDFCI solver addresses
problem~\eqref{eq:optimization} through a coordinate
descent framework incorporating the Gauss--Southwell rule and an exact
line search for step determination.
In what follows, we denote \(f(\vvc)\) as
\(f(\vvc; H)\) whenever the Hamiltonian \(H\) is clear from the context.

At the \(\ell\)-th iteration, the algorithm consists of two steps:
(i) selecting the coordinate to be updated, which has the largest
absolute gradient value,
and (ii) determining the corresponding step size, which
gives the largest descent on the current function value.
These two steps can be expressed as the following
subproblems
\begin{equation}
  \label{eq:choose-i}
    \begin{aligned}
  \ichosen &= \argmax_i \, |(\nabla f(\currvc))_i | \\
  &= \argmax_i \, |4 (H \currvc)_i
  + 4 ({\currvc}^\T \currvc) \currvc_i|,
    \end{aligned}
\end{equation}
and
\begin{equation}
  \label{eq:choose-alpha}
  \begin{aligned}
  \achosen &= \argmin_\eta  \, f(\currvc +
   \eta \vve_{\ichosen}) \\
   &= \argmin_\eta \,  \eta^4 + 4 \currvc_{\ichosen} \eta^3
   + (2 H_{\ichosen, \ichosen} + 4 (\currvc_{\ichosen})^2
   + 2 {\currvc}^\T \currvc ) \eta^2 \\
   & \qquad \qquad \, +(4 (H \currvc)_{\ichosen} + 4 ({\currvc}^\T \currvc) \currvc_{\ichosen}) \eta
   + f(\currvc),
   \end{aligned}
\end{equation}
where the subscript
\(\vvc_i\) denotes the \(i\)-th coordinate of the vector \(\vvc\).
Problem~\eqref{eq:choose-alpha} can be solved by reducing the minimization
of a quartic polynomial to finding the roots of its cubic derivative.
The complete derivation is presented in Appendix~\ref{app:line-search}.
The \(\ichosen\)-th coordinate of \(\vvc\) will be
updated subsequently before moving on to the \((\ell+1)\)-th iteration.

In both steps \eqref{eq:choose-i} and \eqref{eq:choose-alpha},
the vector \(H \currvc\) and the scalar \({\currvc}^\T \currvc\)
are required. To avoid recomputing these quantities,
the vector \(H \vvc\) and two scalars \({\vvc}^\T \vvc\),
\({\vvc}^\T H \vvc\) are stored in memory and incrementally updated
at the end of each iteration. Let \(\vvb\) denote \(H \vvc\),
\(\ccnorm\) and \(\ccquad\) denote  \({\vvc}^\T \vvc\)
and \({\vvc}^\T H \vvc\) respectively.
The update of these variables only uses nonzero entries
of one column of the Hamiltonian matrix:
\begin{equation}
  \label{eq:update-var}
\begin{split}
\nextvc & \gets \vvc^{(\ell)}
          + \achosen \vve_{\ichosen}, \\
\nextvb & \gets \vvb^{(\ell)}
          + \achosen H_{:,\ichosen}, \\
\ccnorm^{(\ell+1)} & \gets \ccnorm^{(\ell)}
                   + 2 \achosen \currvc_{\ichosen} + (\achosen)^2, \\
\ccquad^{(\ell+1)} & \gets \ccquad^{(\ell)}
                   + 2 \achosen \currvb_{\ichosen} + (\achosen)^2
                   H_{\ichosen, \ichosen}.
\end{split}
\end{equation}

During the update of \(\vvb\), a compression scheme can be employed
to restrict the size of the variational space and force the
solution vector \(\vvc\) to converge in a subspace controlled by
a deterministic threshold. The current compression strategy
we use is as follows. Given a fixed threshold \(\tau\),
\(\nextvb_j\) will only be updated if
\(\nextvc_j \ne 0\) or if
\(|\achosen H_{j, \ichosen}| \ge \tau\). This truncation
rule will not affect \(\ccnorm^{(\ell+1)}\) and
\(\ccquad^{(\ell+1)}\) and is both effective and cheap.

Two additional considerations should be mentioned here.
First, when selecting a descent coordinate for
problem~\eqref{eq:choose-i},
an exhaustive search over the full CI space is
computationally infeasible.
Instead, we restrict our search space to \(\Iset(i^{(\ell)})\),
where \(\Iset(i)\) denotes the index set composed of
all indices \(j\) such that \(H_{i,j} \ne 0\).
Second, a modification is introduced in the update of \(\vvb\):
the \(\ichosen\)-th entry in \(\nextvb\) is explicitly recalculated
using the corresponding Hamiltonian entry and coefficients of
\(\vvc\) that are already computed or retrieved,
\begin{equation}
  \label{eq:recalculate-b}
  \nextvb_{\ichosen} = H_{\ichosen, :} \nextvc
  = \sum_{j \in \Iset(\ichosen)} H_{j, \ichosen} \nextvc_j.
\end{equation}
This recalculation improves numerical stability
and ensures the correctness
when entries of \(\vvb\) are truncated
by some threshold \(\tau\).

The complete algorithm for the CDFCI solver~\cite{wangCoordinateDescentFull2019}
is detailed in Algorithm~\ref{alg:cdfci}.

\begin{algorithm}
\caption{CDFCI: Single-Coordinate Descent Method for Ground State}
\label{alg:cdfci}
\begin{algorithmic}[1]
\State Initialize \(\vvc^{(0)}\), \(\vvb^{(0)} = H\vvc^{(0)}\)
and \(i^{(0)}\).
\For{\(\ell = 0,1,2,\dots\) until convergence}
    \State Select coordinate \[\ichosen =
    \argmax_{i \in \Iset(i^{(\ell)})} \,
    |4\currvb_i + 4 \ccnorm^{(\ell)} \currvc_i |.\]
    \State Compute the step size by finding roots of a cubic polynomial
    \( \frac{\diff h(\eta)}{\diff \eta} \)
    (see Appendix~\ref{app:line-search} for details):
     \[ \begin{aligned}
    \achosen & =\argmin_\eta  \, h(\eta)
    \coloneqq f(\currvc + \eta \vve_{\ichosen}) \\
    & = \argmin_\eta \, \eta^4 + 4 \currvc_{\ichosen} \eta^3
   + \left(2 H_{\ichosen, \ichosen} + 4 (\currvc_{\ichosen})^2
   + 2 \ccnorm^{(\ell)} \right) \eta^2 \\
   & \quad + \left(4 \currvb_{\ichosen} +
    4 \ccnorm^{(\ell)}\currvc_{\ichosen}\right) \eta
   + f(\currvc).
     \end{aligned}\]
    \State Update \(\nextvc\) and \(\nextvb\)
\[
\begin{split}
\nextvc & \gets \vvc^{(\ell)}
                           + \achosen \vve_{\ichosen}, \\
\nextvb & \gets \vvb^{(\ell)}
                           + \achosen H_{:,\ichosen}.
\end{split}
\]
    Employ compression on \(\nextvb\) if desired.
    \State Recalculate \(\nextvb_{\ichosen}\)
\[
  \nextvb_{\ichosen} = H_{\ichosen, :} \nextvc
  = \sum_{j \in \Iset(\ichosen)} H_{j, \ichosen} \nextvc_j.
\]
    \State Update \(\ccnorm^{(\ell+1)}\) and \(\ccquad^{(\ell+1)}\)
\[
\begin{split}
  \ccnorm^{(\ell+1)} & \gets \ccnorm^{(\ell)}
                   + 2 \achosen \currvc_{\ichosen} + (\achosen)^2, \\
\ccquad^{(\ell+1)} & \gets \ccquad^{(\ell)}
                   + 2 \achosen \currvb_{\ichosen} + (\achosen)^2
                   H_{\ichosen, \ichosen}.
\end{split}
\]
    \State Report energy estimation as the Rayleigh quotient of
    \(\nextvc\): \(\ccquad^{(\ell+1)}/\ccnorm^{(\ell+1)}\).
\EndFor
\end{algorithmic}
\end{algorithm}

\subsection{Multicoordinate-Descent Methods for Ground State}
\label{sec:mcdfci}
We sketch an extension of Algorithm~\ref{alg:cdfci}
where more than one coordinate can be chosen in each iteration.
It is reported in~\cite{zhangParallelMulticoordinateDescent2025}
that the multicoordinate update strategy
exhibits a similar descent behavior
to standard coordinate descent: updating \(m\) coordinates in one iteration
achieves an energy decrease comparable to
performing \(m\) sequential single-coordinate updates.
The workload per iteration is increased \(m\) times,
and the coordinate updates can be computed concurrently,
making the method more attractive for the parallel processing environment.
In this modified coordinate descent framework,
the update of \(\vvc\) is realized by
\begin{equation}
  \label{eq:update-c}
  \nextvc \gets \gachosen \currvc + \E_{\Ichosen} \vachosen,
\end{equation}
where \(\Ichosen\) denotes the index set of coordinates
chosen, \(|\Ichosen| = m\) and the step size
vector \(\vachosen\) is an extension to the step size scalar
\(\eta\), encoding the step sizes in each direction.
The projection
matrix \(\E_{\Ichosen} \in \mathbb{R}^{\NFCI \times m}\)
is defined as \(\E_{\Ichosen} = [\vve_{i_1^{(\ell+1)}},
\dotsc, \vve_{i_m^{(\ell+1)}}]\) with
\(\Ichosen = \{i_1^{(\ell+1)}, \dotsc, i_m^{(\ell+1)}\}\),
which, in other words, consists of columns from the identity
matrix corresponding to the selected coordinates.
Note that a scaling factor \(\gachosen\) is
inserted into the update formula~\eqref{eq:update-c}
to enable exact line search in a multicoordinate setting.
When \(|\Ichosen| = 1\) and \(\gachosen = 1\),
the original update rule is recovered.

In coordinate selection step, \(m\) coordinates
with largest absolute gradient values are picked out,
\begin{equation}
  \label{eq:choose-I}
  \Ichosen = \Bigl\{ \, i_j^{(\ell+1)}, j = 1, \dotsc, m:
  i_j^{(\ell+1)} = \argmax_{\substack{
    i \in \Iset(\mathcal{I}^{(\ell)}) \\
    i \ne i_1^{(\ell+1)}, \dotsc, i_{j-1}^{(\ell+1)}
   }} \, \bigl|4\currvb_i + 4 \ccnorm^{(\ell)} \currvc_i \bigr| \, \Bigr\}
\end{equation}
where \(\Iset(\mathcal{I}^{(\ell)})\) is the union of
\(\Iset(i)\) for all \(i \in \mathcal{I}^{(\ell)}\).
Moving forward, the optimal step size vector \(\vachosen\)
and scaling factor \(\gachosen\) are determined by
minimizing the function value of the next iterate,
\begin{equation}
  \label{eq:choose-va-ga}
  \begin{aligned}
  \gachosen, \vachosen &=
  \argmin_{\gamma \in \mathbb{R}, \vec{\eta} \in \mathbb{R}^m}
  f(\gamma \currvc + \E_{\Ichosen} \vec{\eta}) \\
  &= \argmin_{\gamma \in \mathbb{R}, \vec{\eta} \in \mathbb{R}^m}
  \, f\left(
  \begin{bmatrix}
     \currvc & \E_{\Ichosen}
  \end{bmatrix}  \begin{bmatrix}
     \gamma \\ \vec{\eta}
  \end{bmatrix}\right). \\
  \end{aligned}
\end{equation}

The matrix
\(\begin{bmatrix}\currvc & \E_{\Ichosen}\end{bmatrix}\)
admits the factorization
\begin{equation}
\begin{bmatrix}
     \currvc & \E_{\Ichosen}
  \end{bmatrix}
= \begin{bmatrix}
     \tilde{\vvc}^{(\ell)} & \E_{\Ichosen}
\end{bmatrix}
\begin{bmatrix}
     \|\currvc - \E_{\Ichosen} \currvc_{\Ichosen}\| & O \\
     \currvc_{\Ichosen} & I_m
\end{bmatrix}
\end{equation}
with
\begin{equation}
\tilde{\vvc}^{(\ell)} =
\begin{cases}
\frac{\currvc - \E_{\Ichosen} \currvc_{\Ichosen}}
{\|\currvc - \E_{\Ichosen} \currvc_{\Ichosen}\|},
& \text{ if } \|\currvc - \E_{\Ichosen} \currvc_{\Ichosen}\|\ne 0,\\
0, & \text{ otherwise}.
\end{cases}
\end{equation}
Let
\(Q^{(\ell+1)} \coloneqq\begin{bmatrix}
     \tilde{\vvc}^{(\ell)} & \E_{\Ichosen}
\end{bmatrix}\), which has orthogonal
columns.
Then \(\nextvc\) can be expressed as a vector
in the span of \(Q^{(\ell+1)}\)
\begin{equation}
  \nextvc = Q^{(\ell+1)} \vec{z}^{(\ell+1)},
\end{equation}
with
\begin{equation}
  \label{eq:def-z}
  \vec{z}^{(\ell+1)} =
  \begin{bmatrix}
     \|\currvc - \E_{\Ichosen} \currvc_{\Ichosen}\| & O \\
     \currvc_{\Ichosen} & I_m
\end{bmatrix}
\begin{bmatrix}
     \gachosen   \\ \vachosen
  \end{bmatrix}.
\end{equation}

Using the orthogonal invariance property
of the Frobenius norm, problem~\eqref{eq:choose-va-ga}
is equivalent to a reduced-size problem
of dimensions \((m+1)\):
\begin{equation}
  \begin{aligned}
  \vec{z}^{(\ell+1)} &=
    \argmin_{\vec{z} \in \mathbb{R}^{m+1}}   \,
f\left(
  Q^{(\ell+1)}
  \vec{z}^{(\ell+1)}
; H \right) \\
&=
  \argmin_{\vec{z} \in \mathbb{R}^{m+1}}   \,
  f\left(
\vec{z}^{(\ell+1)}; (Q^{(\ell+1)})^\T H Q^{(\ell+1)}\right)
\end{aligned}
\end{equation}
and can be solved via seeking the smallest eigenpair of
\((Q^{(\ell+1)})^\T H Q^{(\ell+1)}\). Optimal values of \(\gachosen\)
and \(\vachosen\) are subsequently revealed.

The complete algorithm for multicoordinate descent
method for ground state
computation~\cite{zhangParallelMulticoordinateDescent2025} is detailed in
Algorithm~\ref{alg:mcdfci}.

\begin{algorithm}
\caption{mCDFCI: Multicoordinate Descent Method for Ground State}
\label{alg:mcdfci}
\begin{algorithmic}[1]
\State Initialize \(\vvc^{(0)}\), \(\vvb^{(0)} = H\vvc^{(0)}\)
and \(\I^{(0)}\).
\For{\(\ell = 0,1,2,\dots\) until convergence}
    \State Select coordinates \[  \Ichosen = \Bigl\{ \, i_j^{(\ell+1)}, j = 1, \dotsc, m:
  i_j^{(\ell+1)} = \argmax_{\substack{
    i \in \Iset(\mathcal{I}^{(\ell)}) \\
    i \ne i_1^{(\ell+1)}, \dotsc, i_{j-1}^{(\ell+1)}
   }} \, \bigl|4\currvb_i + 4 \ccnorm^{(\ell)} \currvc_i \bigr| \, \Bigr\}.\]
    \State Compute scaling factor \(\gachosen\)
    and step size vector \(\vachosen\) by
    solving a \((m+1)\)-dimensional eigenvalue problem
\[
    \gachosen, \vachosen =
    \argmin_{\gamma \in \mathbb{R}, \vec{\eta} \in \mathbb{R}^m}   \,
      f\left(
\begin{bmatrix}
     \|\currvc - \E_{\Ichosen} \currvc_{\Ichosen}\| & O \\
     \currvc_{\Ichosen} & I_m
\end{bmatrix}
\begin{bmatrix}
    \gamma   \\ \vec{\eta}
  \end{bmatrix}; (Q^{(\ell+1)})^\T H Q^{(\ell+1)} \right).
\]
    \State Update \(\nextvc\) and \(\nextvb\)
\[
\begin{split}
\nextvc & \gets \gachosen \vvc^{(\ell)}
                      + \E_{\Ichosen} \vachosen, \\
\nextvb & \gets \gachosen \vvb^{(\ell)}
                      + H_{:,\Ichosen} \vachosen.
\end{split}
\]
    Employ compression on \(\vvb\) if desired.
    \State Recalculate \(\nextvb_{\Ichosen}\)
\[
  \nextvb_{i} = H_{i, :} \nextvc
  = \sum_{j \in \Iset(i)} H_{j, i} \nextvc_j,
  \quad \text{for } i \in \Ichosen.
\]
    \State Update \(\ccnorm^{(\ell+1)}\) and \(\ccquad^{(\ell+1)}\)
\[
\begin{split}
\ccnorm^{(\ell+1)} & \gets (\gachosen)^2 \ccnorm^{(\ell)}
                   + 2 \gachosen (\vachosen)^\T \currvc_{\Ichosen}
                   + (\vachosen)^\T \vachosen, \\
\ccquad^{(\ell+1)} & \gets (\gachosen)^2 \ccquad^{(\ell)}
                   + 2 \gachosen (\vachosen)^\T \currvb_{\Ichosen}
                   + (\vachosen)^\T H_{\Ichosen, \Ichosen}
                   \vachosen. \\
\end{split}
\]
    \State Report energy estimation as the Rayleigh quotient of
    \(\nextvc\): \(\ccquad^{(\ell+1)}/\ccnorm^{(\ell+1)}\).
\EndFor
\end{algorithmic}
\end{algorithm}

\subsection{Coordinate-Descent Methods for Excited States}
\label{sec:xcdfci}
Consider problem~\eqref{eq:optimization-K}
when \(K\) low-lying excited states are targeted
in addition to the ground state.
Since now the optimization variable is matrix
\(C \in \mathbb{R}^{\NFCI \times (K+1)}\),
we introduce the following notations:
\(B = HC\) as an extension to the \(\vvb\) vector,
\(G = 4B + 4C(C^\T C)\) representing
the gradient of \(f\), and two
\((K+1) \times (K+1)\) matrices
\(\CCnorm = C^\T C\) and \(\CCquad = C^\T H C\)
as extensions to scalars \(\ccnorm\) and \(\ccquad\).
At each iteration, one row of \(C\) is updated by
\begin{equation}
\nextC \gets \currC + \achosen \vve_{\ichosen}
\currG_{\ichosen, :}.
\end{equation}
The algorithm identifies the next coordinate
and step size by solving two subproblems:
\begin{equation}
  \ichosen = \argmax_i \, \|\currG_{i, :}\|_\infty
\end{equation}
and
\begin{equation}
  \label{eq:choose-eta}
  \achosen = \argmin_\eta f(\currC + \eta
  \vve_{\ichosen}\currG_{\ichosen, :} )
\end{equation}
which again requires minimizing a fourth-order polynomial
with respect to scalar \(\eta\).
Computation details are provided in
Appendix~\ref{app:line-search}.

The minimizers of~\eqref{eq:optimization-K} only
give eigenspaces, but not orthogonal eigenvectors.
To retrieve eigenvectors, we need another
post-processing step of solving the following
general eigenvalue problem of size \((K+1) \times (K+1)\):
\begin{equation}
  \CCquad U = \CCnorm U \Gamma
\end{equation}
for \(U\) being eigenvectors and \(\Gamma\) being the
eigenvalue matrix.

The complete coordinate descent algorithm for excited states~\cite{wangCoordinateDescentFull2023}
is summarized in
Algorithm~\ref{alg:xcdfci}.

\begin{algorithm}
\caption{xCDFCI: Coordinate Descent Method for Excited States}
\label{alg:xcdfci}
\begin{algorithmic}[1]
\State Initialize \(C^{(0)}\), \(B^{(0)} = HC^{(0)}\)
and \(i^{(0)}\).
\For{\(\ell = 0,1,2,\dots\) until convergence}
    \State Select coordinate
    \[\ichosen = \argmax_{i\in \Iset(i^{(\ell)})} \, \|\currG_{i, :}\|_\infty\]
    where \(\currG = 4\currB + 4\currC \CCnorm^{(\ell)}\).
    \State Compute the step size by
    solving the roots of a cubic polynomial
    \( \frac{\diff h(\eta)}{\diff \eta} \)
    (see Appendix~\ref{app:line-search} for details):
     \[
    \achosen =\argmin_\eta  \, h(\eta)
    \coloneqq f(\currC + \eta
  \vve_{\ichosen}\currG_{\ichosen, :} ).\]
    \State Update \(\nextC\) and \(\nextB\)
\[
\begin{split}
\nextC       &\gets \currC + \achosen \vve_{\ichosen}
\currG_{\ichosen, :}, \\
\nextB       &\gets \currB + \achosen H_{:,\ichosen}
\currG_{\ichosen, :}. \\
\end{split}
\]
    Employ compression on rows of \(\nextB\) if desired.
    \State Recalculate \(\nextB_{\ichosen}\)
\[
  \nextB_{\ichosen}
  = \sum_{j \in \Iset(\ichosen)} H_{j, \ichosen} \nextC_{j,:}.
\]
    \State Update \(\CCnorm^{(\ell+1)}\) and \(\CCquad^{(\ell+1)}\)
\[
\begin{split}
\CCnorm^{(\ell+1)} & \gets \CCnorm^{(\ell)}
                   + \achosen
           \left({\currC_{\ichosen, :}}^\T \currG_{\ichosen, :}
           +{\currG_{\ichosen, :}}^\T \currC_{\ichosen, :}\right) \\
                   &\quad + (\achosen)^2 \|\currG_{\ichosen, :}\|^2,\\
\CCquad^{(\ell+1)} & \gets \CCquad^{(\ell)}
                   + \achosen
           \left({\currB_{\ichosen, :}}^\T \currG_{\ichosen, :}
           +{\currG_{\ichosen, :}}^\T \currB_{\ichosen, :}\right) \\
                   &\quad + (\achosen)^2 H_{\ichosen, \ichosen}
                   \|\currG_{\ichosen, :}\|^2.
\end{split}
\]
    \State Report energy estimation after solving
    the general eigenvalue problem of matrix pencil
    \((\CCquad^{(\ell+1)}, \CCnorm^{(\ell+1)})\).
\EndFor
\end{algorithmic}
\end{algorithm}

\subsection{Optimal Orbital Selections}
\label{sec:optorbfci}
In scenarios where the memory budget is limited,
it is often beneficial to optimize the one-particle
basis functions to achieve a more compact representation
of the wavefunction. This approach is particularly useful
when the full configuration interaction (FCI) space is too large
to be fully explored.
Given \(\morb\) one-particle basis functions
\(\{\psi_1, \dots, \psi_\morb\}\),
the goal is to find an orthogonal matrix
\(U \in \mathbb{R}^{\morb \times \norb}, U^\T U = I_\norb\)
that transforms the basis functions
\begin{equation}
    (\phi_1, \dots, \phi_\norb) = (\psi_1, \dots, \psi_\morb)  U,
\end{equation}
such that the transformed basis functions
\(\{\phi_1, \dots, \phi_\norb\}\) minimize the ground-state
energy within a smaller FCI space.

To solve this problem, we employ a two-level optimization strategy,
corresponding to the two levels of variables: the rotation matrix
\(U\) and the wavefunction coefficients \(\vvc\).
The outer level that focuses on optimizing the one-particle
functions formulates the following optimization problem:
\begin{equation}
  \label{eq:subopt}
    \min_{U \in \mathbb{R}^{\morb \times \norb}, U^\T U = I_\norb} \,
    E_0(U) \coloneqq \sum_{p',q' = 1}^\norb \sum_{p,q = 1}^\morb h_{pq} U_{pp'}
        U_{qq'} \oneD^{p'}_{q'}  + \sum_{p',q',r',s' = 1}^\norb \sum_{p,q,r,s
        = 1}^\morb v_{pqrs} U_{pp'} U_{qq'} U_{rr'} U_{ss'}
        \twoD^{p'q'}_{r's'}
\end{equation}
where $\oneD^{p'}_{q'} = \ev**{\op{a}_{p'}^\dagger \op{a}_{q'}}{\Phi}$
and $\twoD^{p'q'}_{r's'} = \ev**{\op{a}_{p'}^\dagger
\op{a}_{q'}^\dagger \op{a}_{s'} \op{a}_{r'}}{\Phi}$ are the
one-body and two-body reduced density matrices
(1RDM and 2RDM) associated with the wavefunction \(\Phi\),
which is represented in the \(\norb\)-particle Hilbert space
spanned by the transformed single-particle basis functions
\(\{\phi_1, \dots, \phi_\norb\}\).
The objective function is a fourth-order polynomial of $U$ and
can be optimized
using gradient-based projection methods or other suitable optimization
techniques on Stiefel manifolds.
In the current implementation, projected gradient descent with
alternating Barzilai--Borwein (BB) step size
is employed to solve problem~\eqref{eq:subopt}.
On the other hand,
the inner level is just solving the FCI problem
given a fixed set of one-particle basis functions,
and can be efficiently handled by the coordinate descent methods
described in the previous sections.

The full algorithm for optimal orbital FCI
(OptOrbFCI)~\cite{liOptimalOrbitalSelection2020}
is summarized in Algorithm~\ref{alg:optorbfci}.

\begin{algorithm}
\caption{OptOrbFCI: Optimal Orbital Selection with Coordinate Descent FCI}
\label{alg:optorbfci}
\begin{algorithmic}[1]
\State Initialize one-particle basis functions
\(\{\phi_i^{(0)}\}_{i=1}^\norb\)
from a larger set \(\{\psi_i\}_{i=1}^\morb\).
\For{\(\ell = 0,1,2,\dots\) until convergence}
    \State Compute one- and two-electron integrals
    \(\{h_{pq}^{(\ell)}\}_{p,q=1}^\norb\) and
    \(\{v_{pqrs}^{(\ell)}\}_{p,q,r,s=1}^\norb\) in the current basis
    \(\{\phi_i^{(\ell)}\}_{i=1}^\norb\).
    \State Solve the FCI problem in the current basis
    using coordinate descent methods to obtain
    ground-state energy \(E_0^{(\ell)}\)
    and wavefunction coefficients \(\vvc^{(\ell)}\).
    \State Compute the 1RDM and 2RDM from the
    ground-state wavefunction.
    \State Solve the orthonormal constrained polynomial
        \eqref{eq:subopt} via projection method with alternating BB
        step size as
\begin{equation} \label{eq:projBB}
    U^{(\ell+1)} = \orth{U^{(\ell)} - \tau_k \grad_{U^{(\ell)}} P_4(U^
    {(\ell)})},
\end{equation}
        and obtain \(U^{(\ell+1)}\). Here, \(\orth{X}\) denotes the
        orthogonalization of matrix \(X\).
    \State Update one-particle basis functions
    \((\phi_1^{(\ell+1)}, \dots, \phi_\norb^{(\ell+1)}) =
    (\psi_1, \dots, \psi_\morb) U^{(\ell+1)}\).
\EndFor
\end{algorithmic}
\end{algorithm}



\section{Overview of the CDFCI Software Package}
\label{sec:software}
CDFCI is implemented in C++ and designed with a modular architecture.
It leverages advanced features of the C++17 standard and uses the \texttt{Eigen}
library~\cite{eigen} for small and dense linear algebra operations.
Section~\ref{sec:data-structure} presents three major classes that 
constitute the core of the implementation: the \texttt{Hamiltonian} class, 
the \texttt{WaveFunction} class, and the \texttt{Solver} class. We discuss the 
parallelization support in Section~\ref{sec:parallelization-support} and 
analyze the computational complexity and memory usage in 
Section~\ref{sec:computational-complexity}.
For end users, Section~\ref{sec:example-usage} provides a usage example
demonstrating how to set up and run a CDFCI calculation.
For developers interested in extending CDFCI,
Section~\ref{sec:extending-cdfci} describes the main extension points
for implementing custom Hamiltonians, replacing the wavefunction storage
backend, designing coordinate strategies, and integrating new solvers.

\subsection{Core Abstractions and Data Structures}
\label{sec:data-structure}
\subsubsection{The \texttt{Hamiltonian} Class}
\label{sec:hamiltonian}
The Hamiltonian class is responsible for representing
the fermionic Hamiltonian in the second-quantized form.
It provides methods for constructing the Hamiltonian,
as well as accessing its matrix elements efficiently.

The abstract class \texttt{Hamiltonian<N>} uses the same template
parameter \texttt{N} as \texttt{Determinant<N>}. Specifically,
\texttt{N} is the number of \texttt{size\_t} words used to store
the occupation bit string of a Slater determinant.
Note that the Hamiltonian matrix is never generated
or stored explicitly. Instead, several access methods
are provided to retrieve matrix elements or columns on-the-fly:
\begin{itemize}
  \item \texttt{get\_entry(\(D_1, D_2\))}: returns the matrix element
  \(H_{D_1, D_2}\),
  according to the Slater--Condon rules~\eqref{eq:entry}.
  \item \texttt{get\_diagonal(\(D\))}: returns the diagonal element
  \(H_{D, D}\).
  \item \texttt{get\_column(\(D\))}: returns all nonzero elements
  in the column corresponding to the determinant \(D\), as an unordered
  collection of \((D', H_{D', D})\) tuples.
\end{itemize}
This interface is intentionally designed to match the needs of
coordinate-descent updates, as the algorithm repeatedly requires
column access to \(H\) with respect to the current determinant.

Currently, two derived classes of \texttt{Hamiltonian<N>} are provided:
\texttt{HamiltonianMolecule<N>} for molecular Hamiltonians,
and \texttt{HamiltonianHubbardK<N>} which implements a \(D\)-dimensional
Hubbard model with periodic boundary conditions
over an orthogonal lattice.
For electronic structure calculations, the Hamiltonian
is read from FCIDUMP files, which are a standard format
for storing one-body and two-body integrals in quantum chemistry.
For lattice models, such as the Hubbard model,
the Hamiltonian can be constructed directly from
the model parameters.
Refer to Section~\ref{sec:implement-hamiltonian} for details on implementing a custom Hamiltonian.

\subsubsection{The \texttt{WaveFunction} Class}
\label{sec:wavefunction}
In the \texttt{WaveFunction} class, the sparse vectors
\(\vvc\) and \(\vvb\) (or rowwise sparse matrices \(C\) and \(B\)
in the multi-state case) are represented by a key-value mapping,
where each key denotes a Slater determinant and each value
contains the associated coefficients.
For each \texttt{key\_type} \(D\),
\[
  \texttt{mapped\_type} =
  \begin{cases}
    (\vvc_D, \vvb_D), & \text{if } \texttt{NSTATES}=1,\\
    (C_{D,:}, B_{D,:}), & \text{if } \texttt{NSTATES}>1,
  \end{cases}
\]
where \texttt{NSTATES} is a template parameter
controlling the number of states.

The class template \(\texttt{WaveFunction<Container, NSTATES>}\)
receives its storage backend as a template parameter. The current
single-threaded implementation uses a robin-hood hash
map~\cite{robinhood}, whereas the OpenMP implementation uses a concurrent
cuckoo hash map~\cite{Fan2013,Li2014,libcuckoo}. Both are accessed through
the adapter class \texttt{Container} that exposes the common operations 
required by \texttt{WaveFunction}; see Section~\ref{sec:storage-backend} for how to add another backend.
The solver monitors the number of stored entries against a configured capacity
limit and reports an explicit overflow error when that limit is exceeded,
preventing uncontrolled growth beyond the intended memory budget.

In addition to \(\vvc\) and \(\vvb\) (or \(C\) and \(B\)),
the class also maintains the vector norm
\(\ccnorm = \vvc^\T \vvc \) (or \(\CCnorm = C^\T C\)),
the quadratic form
\(\ccquad = \vvc^\T \vvb\) (or \(\CCquad = C^\T B\)),
and the scaling factor \(\gamma\) introduced in
Section~\ref{sec:cdfci}-\ref{sec:xcdfci}.
These variables are accumulated in quadruple precision to suppress
the propagation of rounding errors during millions of iterative updates,
thereby preserving the numerical stability of the energy estimate.

All the above variables are updated in each iteration through
the function \texttt{update\_coordinate(\allowbreak det\_picked,\allowbreak h,\allowbreak sub\_xz)}.
The exact update rule depends on the chosen algorithm.
The truncation threshold \(\tau\), which controls
the sparsity of the wavefunction, is provided
by the user via the parameter \texttt{z\_threshold}.
Finally, the class provides the method \texttt{get\_variational\_energy()}
to compute the Rayleigh quotient or solve the generalized eigenvalue problem
for energy estimation.
For the ground state (\(\texttt{NSTATES}=1\)), the Rayleigh quotient is
\[
  E_0 = \frac{\ccquad}{\ccnorm} = \frac{\vvc^\T \vvb}{\vvc^\T \vvc}.
\]
For \(\texttt{NSTATES}=(K+1)>1\), the method forms the generalized eigenproblem
\(
  \CCquad u = \CCnorm u \lambda.
\)
The \(K+1\) smallest eigenvalues are
then computed using \texttt{GeneralizedSelfAdjointEigenSolver}
from the \texttt{Eigen}~\cite{eigen} library, and the eigenvalues are
reported in ascending order.

\subsubsection{Solvers for Ground and Excited States}
\label{sec:solvers}
We develop a unified solver framework capable of handling both
ground and excited states within the same coordinate-descent
paradigm. The generic \texttt{Solver<H,W>} class
accepts user-defined Hamiltonian \texttt{H} and wavefunction
\texttt{W} as template parameters, as well as
pluggable \texttt{CoordinatePick} and
\texttt{CoordinateUpdate} strategies.

The \texttt{Solver<H,W>::solve} routine performs the following
operations in each iteration:
\begin{enumerate}
  \item select one or multiple determinants using \texttt{CoordinatePick};
  \item compute step size(s) and the scaling factor by \texttt{CoordinateUpdate};
  \item call \texttt{W::update\_coordinate} to update \(\vvc\) and \(\vvb\)
  (or \(C\) and \(B\));
  \item call \texttt{W::get\_variational\_energy} to estimate energy
  and check convergence.
\end{enumerate}

In the last step, the stopping criterion can be user-defined.
For the single-coordinate ground state algorithm,
we adopt a damped accumulator
\[
  d^{(\ell+1)} \gets \theta d^{(\ell)} + (1-\theta)\|\eta^{(\ell+1)}\|_2,\qquad \theta\in[0,1),
\]
with \(\theta=\) \texttt{stopping\_dx\_damping\_factor} and
\(t=\) \texttt{stopping\_dx\_threshold}. The run stops when
\(d^{(\ell+1)}<t\).
The same idea is naturally extended to multicoordinate or multi-state variants.

The iteration loop employs a configurable \(\texttt{report\_interval}\)
to balance monitoring frequency and computational overhead.
At each report, diagnostic quantities such as iteration count,
variational energies, the damped accumulator,
sparsity statistics, and elapsed wall time are recorded.
The solver further supports checkpointing for reliable termination
and restart from stored wavefunctions, facilitating long-duration simulations.

The solver variants introduced in Section~\ref{sec:methodology}
are implemented as derived classes of the generic template \texttt{Solver<H, W>}.
The subsequent paragraphs describe the
design of each solver.

\paragraph{Ground State CDFCI}
The \texttt{CDFCISolver} realizes the ground state algorithm by
specializing the \texttt{CoordinatePick} and
\texttt{CoordinateUpdate} strategies.
When \texttt{num\_coordinates} is one, \texttt{CDFCISolver} selects
\texttt{gcd\_grad} and \texttt{ls}, reproducing the original CDFCI method 
(Alg.~\ref{alg:cdfci}).
When \texttt{num\_coordinates} is greater than one, the same class instead
selects \texttt{block\_gcd\_grad} and \texttt{eig}, which implement the
block coordinate selection and local eigenvalue update of mCDFCI 
(Alg.~\ref{alg:mcdfci}).
Initialization can use either a single reference determinant,
for example the Hartree--Fock state obtained from 
\texttt{H::get\_hartree\_fock},
or user-provided configurations.

\paragraph{Excited-state xCDFCI}
The \(\texttt{XCDFCISolver}\) operates with \(\texttt{NSTATES}= (K+1) >1\)
and initializes \((K+1)\) starting determinants \(\{D_i^{(0)}\}_{i=1}^{K+1}\)
by selecting the Hartree--Fock determinant together with low-diagonal-energy
determinants connected to it,
\[
D_i^{(0)} = \argmin_{
\substack {
  j \notin \{D_1^{(0)}, \dotsc, D_{i-1}^{(0)}\} \\
  j \in \Iset(i_{\text{HF}})
}} \, H_{j,j}, \quad i = 1, \dotsc, K+1.
\]
The initial coefficient matrix \(C^{(0)}\) is constructed as
\[
C^{(0)}_{i,j} =
\begin{cases}
  1, &i=D_j^{(0)}, \\
  0, &\text{otherwise},
\end{cases}
\]
and \(B^{(0)} = HC^{(0)}\).
The subsequent iterations follow the procedure outlined in
Alg.~\ref{alg:xcdfci}.

\paragraph{Orbital-optimized OptOrbFCI}
The OptOrbFCI method (Alg.~\ref{alg:optorbfci}) is implemented at the
program-orchestration level, outside the
\texttt{Solver<H,W>} inheritance hierarchy. The \texttt{OptOrbFCIProgram}
maintains the rectangular orbital-transformation matrix and alternates between
two components. First, it constructs the Hamiltonian in the current compressed
orbital basis and invokes a standard \texttt{CDFCIProgram} as the
inner CI solver. It then obtains the one- and two-particle reduced density
matrices from the resulting wavefunction and passes them to the
\texttt{Optimizer}, which updates the orbital transformation using projected
gradient steps with alternating Barzilai--Borwein step sizes. This separation
allows the inner calculation to reuse the same \texttt{Hamiltonian},
\texttt{WaveFunction}, and \texttt{Solver} components as an ordinary CDFCI
calculation, while \texttt{OptOrbFCIProgram}\allowbreak\texttt{Driver} manages the outer stopping
criterion and orbital-compression options.

\subsection{Parallelization Support}
\label{sec:parallelization-support}

As discussed in Section~\ref{sec:computational-complexity}, updating the
wavefunction after a coordinate step is typically the dominant computational
kernel. For selected increments \(\{\Delta \vvc_D\}\), this operation updates
\(\vvb=H\vvc\), before optional compression, by accumulating the
corresponding Hamiltonian columns,
\[
  \vvb \gets \vvb + \sum_D \Delta \vvc_D H_{:,D}.
\]
The implementation parallelizes this matrix-free update with OpenMP at two
complementary levels. When several coordinates are selected, as in mCDFCI,
their Hamiltonian columns are assigned to threads using dynamic scheduling.
Within each selected column, construction is further decomposed into one task
for the diagonal and single-excitation contributions and multiple tasks for
the double-excitation contributions. The latter are partitioned by disjoint
ranges of occupied-orbital pairs and generated through
\texttt{H::get\_column\_parallel\_part}. The current implementation creates
\(\nelec/2\) double-excitation tasks per column, so that their individual
workloads are comparable to that of the diagonal-and-singles task.

All tasks belong to the same OpenMP task group.
Each task constructs a column fragment and applies its contribution in a
thread-local \texttt{sub\_xz\_parallel} buffer. Only the reduction of these
buffers, including the partial update to the selected entry of \(\vvb\), is
performed in an OpenMP critical section. The resulting \texttt{sub\_xz}
contains the determinants affected by the current step and is reused during
coordinate selection in the next iteration.

\subsection{Computational Complexity}
\label{sec:computational-complexity}

Given a determinant \(D\), \texttt{get\_column(D)} returns
\(\{(D,H_{D,D})\}\cup\{(D',H_{D',D})\}\)
over all single and double excitations
from \(D\). Denote \(\norb\) the number of spin--orbitals
and \(\nelec\) the number of electrons.
The diagonal \(H_{D,D}\) is computed in \(\bigO(\nelec^2)\) time
using the diagonal cache. Single excitations \((i\to a)\) are generated
in \(\bigO(\nelec(\norb-\nelec))\) time by scanning occupied \(i\) and
consulting the single-excitation structure. Each matrix element
equals \(h_{ia} + \sum_k \langle ik\Vert ak\rangle\) up to the fermionic
parity sign. Double excitations \((i,j)\to(a,b)\) are generated in
\(\bigO(\nelec^2(\norb-\nelec)^2)\) time by scanning ordered pairs of occupied
\((i,j)\) and consulting the double-excitation structure. 
Each matrix element equals
\(\langle ij\Vert ab\rangle\) up to the fermionic parity sign.
To summarize, the time complexity of \texttt{get\_column(D)} is
\(\bigO(\nelec^2\norb^2)\) in the worst case.

Let \(\NH = \max_i |\Iset(i)|\) denote the
maximum number of nonzero entries in columns of the Hamiltonian
matrix.
For the single-coordinate descent ground state solver CDFCI,
the determinant selection step involves looping over \(\NH\) candidates,
resulting in a computational cost of \(\bigO(\NH)\)
with a small prefactor.
The subsequent line search and update of \(\vvc\) incur only
negligible \(\bigO(1)\) cost.
Updating \(\vvb\) requires evaluating \(\bigO(\NH)\)
Hamiltonian entries and accessing
\(\NH\) entries of \(\vvb\) and \(\vvc\),
leading to an overall \(\bigO(\NH)\)
complexity. The prefactor here is determined by the
per entry evaluation cost of the Hamiltonian,
plus the cost of accessing the underlying data structure of the wavefunction.
Consequently, the leading order of complexity per iteration is \(\bigO(\NH)\).

For the multicoordinate descent variant mCDFCI,
assume that \(m\) coordinates are selected in each iteration.
In the coordinate selection step,
we maintain a size-\(m\) min-heap
during a linear scan over all
\(m\NH\) candidates to track the top
\(m\) coordinates with the largest
gradient magnitudes, combined with
hash-based deduplication,
reducing the complexity from the
naive \(\bigO(m^2\NH)\) to
\(\bigO(m \log m \NH)\).
The line search step is independent of \(\NH\)
and scales with \(m^3\), thus the total serial cost per iteration is
\(\mathcal O(m N_H \log m + m^3).\)
Although the coordinate-selection term has an additional
\(\log m\) factor, updating \(b\) is typically the dominant kernel
in practice because Hamiltonian element evaluation and hash table
access have substantially larger prefactors. With \(p\) OpenMP
threads, the parallelizable part of the update costs approximately
\(\mathcal O(mN_H/p)\), subject to synchronization and load-balancing
overheads.

We use the same notation \(\NH\)
to analyze the computational cost for xCDFCI solver.
In the determinant selection step,
each iteration loops over \(\NH\) candidates
and evaluates the norm of the corresponding
gradient row.
Since this operation involves a
multiplication with a \((K+1) \times (K+1)\) matrix,
the total cost of this step is \(\bigO(\NH (K+1)^2)\).
The line search step is independent of \(\NH\) and
scales with \((K+1)^2\), which is
negligible in practice because typically \(\NH \gg (K+1)^2\).
Updating the coefficient matrices \(C\) and \(B\)
requires accessing \(\bigO(\NH)\) rows and updating
each of the \((K+1)\) states, with an average
constant-time cost per hash table lookup,
resulting in a total complexity of \(\bigO(\NH (K+1))\).
Therefore, the overall leading-order complexity of one
iteration of xCDFCI is bounded by \(\bigO(\NH (K+1)^2)\).

\paragraph{Memory usage.}

In the current implementation, the dominant persistent allocation for
large selected spaces is the hash table used by \texttt{WaveFunction}.
Let \(N_{\mathrm{cap}}\) denote the allocated number of hash table slots and let
\(w=\lceil \norb/64\rceil\) be the number of 64-bit words used to encode one
determinant. Each CDFCI or mCDFCI entry contains an \(8w\)-byte determinant
key and two double-precision values, \(x_i\) and \(z_i\), giving a raw payload
of \(8w+16\) bytes per slot. For xCDFCI with \(K+1\) states, the corresponding
payload is \(8w+16(K+1)\) bytes because both \(x_i\) and \(z_i\) are stored for
every state. Thus, excluding container metadata, the allocated wavefunction
payload is
\begin{equation}
M_{\mathrm{wf}}
=
N_{\mathrm{cap}}
\begin{cases}
8w+16, & \text{CDFCI and mCDFCI},\\
8w+16(K+1), & \text{xCDFCI}.
\end{cases}
\end{equation}
CDFCI limits the number of stored determinants to
\(N_{\max}=\alpha N_{\mathrm{cap}}\), where \(\alpha=0.79\) is the default
occupancy factor used by the program. The actual allocation is slightly larger
than the raw payload because \texttt{libcuckoo} stores entries in fixed-size
buckets with additional occupancy and partial-hash metadata, together with a
small lock array. Other persistent allocations, such as Hamiltonian integral
data, are problem dependent and are typically smaller than the wavefunction
table for the large selected spaces considered here.

\subsection{Example Usage}
\label{sec:example-usage}
This section shows minimal, end-to-end examples for molecular
Hamiltonians, using the ground state CDFCI solver.
We illustrate the basic input format, compilation, and execution workflow.
Examples assume a \texttt{C++17} compiler and optional OpenMP support for shared-memory
parallelism. More examples and a detailed \texttt{README} are provided in the
\texttt{examples/} directory of the source code.

\subsubsection{Input}
The JSON input schema used by CDFCI consists of three blocks:
\texttt{hamiltonian}, \texttt{solver}, and optional global settings such as
\texttt{max\_memory}. Below is a simple input configuration for a
molecular Hamiltonian:

\begin{verbatim}
# input.json
{
  "hamiltonian": {
    "type": "molecule",
    "molecule": { "fcidump_path": "h2o_sto3g.FCIDUMP", "threshold": 0.0}
  },
  "solver":{
      "type": "cdfci",
      "cdfci": {
          "num_iterations": 150000,
          "report_interval": 10000,
          "z_threshold": 0,
          "z_threshold_search": false
      }
  },
  "max_memory": 0.05
}
\end{verbatim}

For chemical molecules, the solver reads a standard \texttt{FCIDUMP} file
containing one- and two-electron integrals,
and a truncation threshold can be set to neglect small integral values.
It then automatically constructs the
Hartree--Fock (HF) reference determinant if not provided, and begins coordinate
updates from the HF state.
Typical parameters under the \texttt{solver/cdfci} block include:
\texttt{num\_iterations} and \texttt{report\_interval}, which control
the iteration loop and output frequency;
\texttt{z\_threshold}, which specifies truncation tolerance \(\tau\) for
the vector \(\vvb\);
and \texttt{z\_threshold\_search}, which can be used to automatically
adjust compression to fit available memory.
The parameter \texttt{max\_memory} limits the in-memory wavefunction size (in GiB).

\subsubsection{Compile and Run}
In a Linux environment, the program is compiled using a Makefile
and run as a binary.

\begin{verbatim}
$ make cdfci
$ bin/cdfci input.json        # single-threaded version
$ bin/cdfci_omp input.json    # multi-threaded version
\end{verbatim}

Compilation produces several solver binaries (e.g., \texttt{cdfci},
\texttt{cdfci\_omp}, \texttt{xcdfci}), located in \texttt{bin/}.
Execution reads the JSON file, loads the Hamiltonian, and
iterates until convergence or the specified iteration limit.
When OpenMP is enabled, the code automatically parallelizes Hamiltonian column
construction and coordinate updates.

\subsubsection{Expected Console Output}
The following shows the expected output of the \ch{H2O}/STO-3G example,
omitting the program header information (e.g., build, machine, input).

\begin{verbatim}
...
CDFCI calculation
-----------------
Reference determinant occupied spin-orbitals:
0  1  2  3  12  13  16  17
Reference energy: -75.5854987695

Iteration  Energy          dx         |x|_0  |z|_0  |H_i|_0  Time
10000     -75.7160281389  5.8949e-04   9109   58105      321  0.29
20000     -75.7160958597  9.9561e-05  17604   60811      409  0.50
30000     -75.7161047048  1.2322e-04  25284   61354      397  0.71
...
Final FCI Energy:           -75.7161071527957006
\end{verbatim}

The central part of the output is a tabulated progress report printed every
\texttt{report\_interval} iterations.
Each row lists the iteration number, current variational energy (in Hartree),
step size \texttt{dx}, the number of nonzero coefficients in the current
wavefunction (\texttt{|x|\_0}) and residual (\texttt{|z|\_0}), the number of nonzero Hamiltonian
entries in selected column(s), and elapsed wall time.
The final energy should converge near \(-75.716107\) Hartree for the
\ch{H2O}/STO-3G example, consistent with reference FCI values.

Additional examples, including OpenMP, multicoordinate, and excited-state
variants, are available in the \texttt{examples/} directory.

\subsubsection{Python Interface}
The C++ implementation is exposed through a \texttt{pybind11} extension and a
high-level Python module named \texttt{cdfci}. The Python API accepts the same
option schema as the C++ JSON interface and provides entry points for CDFCI,
xCDFCI, and OptOrbFCI. For example, a calculation from an existing FCIDUMP file
can be run as follows:
\begin{verbatim}
import cdfci

option = {
    "hamiltonian": {
        "type": "molecule",
        "molecule": {"fcidump_path": "h2o_sto3g.FCIDUMP"},
    },
    "solver": {
        "type": "cdfci",
        "cdfci": {
            "num_iterations": 150000,
            "report_interval": 10000,
        },
    },
    "max_memory": 1.0,
}
res = cdfci.run_cdfci(option)
print(res.energy)
\end{verbatim}

The module integrates with PySCF at two levels. Given a converged PySCF SCF
object, \texttt{run\_from\_pyscf\_scf} transforms the one- and two-electron
integrals to the molecular-orbital basis and passes restricted-Hartree--Fock
integrals directly to CDFCI in memory, avoiding an intermediate FCIDUMP file.
A minimal SCF-to-CDFCI workflow is:

\begin{verbatim}
import cdfci
from pyscf import gto, scf

mol = gto.M(
    atom="H 0 0 0; H 0 0 0.74",
    basis="sto-3g",
    spin=0,
)
mf = scf.RHF(mol).run()

option = {
    "solver": {
        "type": "cdfci",
        "cdfci": {
            "num_iterations": 50000,
            "report_interval": 5000,
        },
    },
    "max_memory": 1.0,
}
res = cdfci.run_from_pyscf_scf(mf, option=option)
print(res.energy)
\end{verbatim}

For active-space workflows, \texttt{CDFCIPySCFFCISolver} implements the
\texttt{kernel}, \texttt{make\_rdm1}, and \texttt{make\_rdm12} methods expected
by PySCF. It can therefore replace the default FCI solver in a PySCF CASCI or
CASSCF object:

\begin{verbatim}
from pyscf import mcscf

mc = mcscf.CASCI(mf, ncas=2, nelecas=2)
mc.fcisolver = cdfci.as_pyscf_fcisolver(option=option)
e_tot = mc.kernel()[0]
print(e_tot)
\end{verbatim}

The adapter computes spin-summed one- and two-particle reduced density matrices
in CDFCI and returns them through the PySCF solver interface, enabling PySCF to
perform the associated active-space energy evaluation and orbital-optimization
steps. Build instructions and advanced examples are provided in the Python
interface user manual distributed with the source code.

\subsection{Extending CDFCI}
\label{sec:extending-cdfci}

CDFCI separates the physical model, sparse wavefunction storage, algorithmic
strategies, and program-level control flow into distinct software components.
This separation permits extensions at different levels without requiring
changes throughout the codebase. In the following, we describe how to add a
custom Hamiltonian, replace the wavefunction storage backend, introduce new
coordinate selection/update strategies, and integrate a solver whose
iteration may not conform to the existing coordinate-descent workflow.

\subsubsection{Implementing a Custom Hamiltonian}
\label{sec:implement-hamiltonian}

New physical models are added by deriving a class from
\texttt{Hamiltonian<N>}, where \texttt{N} is the number of
\texttt{size\_t} words used to represent a determinant. A derived class must
provide the matrix-free operations used by the solver, including
\texttt{get\_entry}, \texttt{get\_diagonal}, and the serial and parallel
column-construction interfaces. It must also provide the model-specific
reference determinant (counterpart of the Hartree--Fock determinant)
and the routines that generate diagonal, single-, and
double-excitation contributions. These operations are sufficient for the
generic solver because no component requires an explicitly assembled
Hamiltonian matrix.

The existing \texttt{HamiltonianMolecule<N>} and
\texttt{HamiltonianHubbardK<N>} implementations provide examples for an
integral-driven Hamiltonian and a parameterized lattice Hamiltonian,
respectively.

\subsubsection{Replacing the Wavefunction Storage Backend}
\label{sec:storage-backend}

The main sparse wavefunction container is selected at compile time through the
\texttt{Container} template parameter of
\texttt{WaveFunction<Container,NSTATES>}. Rather than depending directly on a
particular storage library, \texttt{WaveFunction} accesses it through a thin
adapter that presents a uniform container abstraction. The current
implementation provides \texttt{ContainerRobinhood} for the serial build and
\texttt{ContainerCuckoo} for the OpenMP build, illustrating how storage
implementations with different native interfaces can support the same
wavefunction algorithms.

The essential distinction between the two cases is their concurrency
semantics. A serial backend requires only ordinary associative updates, whereas
an OpenMP backend must support atomic insertion and modification because
multiple tasks may update the same determinant concurrently. Replacing the
backend therefore consists of supplying an adapter with the appropriate serial
or concurrent semantics and selecting it as the program's
\texttt{container\_type}; the higher-level \texttt{WaveFunction} and solver
logic remain unchanged.

\subsubsection{Designing a New Coordinate Selection/Update Strategy}

Coordinate selection and coordinate update are independent strategy
interfaces. A new selection rule derives from \texttt{CoordinatePick<W>} and
implements its function-call operator, which receives the current
\texttt{WaveFunction} and a list of candidate determinants and returns one or
more selected determinants together with the coefficients.
Similarly, a new update rule
derives from \texttt{CoordinateUpdate<H,W>} and computes the coefficient
increments for the selected coordinates and the scaling factor. 
The current implementations include
\texttt{CoordinatePickGcdGrad}, \texttt{CoordinatePick}\allowbreak\texttt{BlockGcdGrad},
\texttt{CoordinateUpdateLS}, and \texttt{CoordinateUpdateEig}.

Each new strategy must be registered under a string identifier in the
corresponding \texttt{init} factory. Once registered, the strategy can be
composed with the generic
iteration machinery by a solver class, without duplicating wavefunction
updates, energy estimation, convergence checks, or reporting.

\subsubsection{Integrating a New Solver}

The strategy extension described above applies when a new method retains the
iteration pattern implemented by \texttt{Solver<H,W>::solve}: select
coordinates, compute their increments, update the wavefunction, and estimate
the energy. To implement a solver with a different iteration structure, 
it must be integrated at the program layer.
The developer first implements an algorithm class with its own
\texttt{solve} or \texttt{run} routine. This class may still reuse the
\texttt{Hamiltonian} interface for matrix-free element or column access, the
determinant representation, the existing sparse \texttt{WaveFunction}, and the
shared-memory kernels. It should expose its final energies and iteration 
statistics through the common \texttt{Result} type. 

The algorithm is then wrapped in a dedicated program class, analogous to
\texttt{CDFCIProgram} or \texttt{XCDFCIProgram}. The program constructs the
Hamiltonian through \texttt{Hamiltonian<N>::init}, selects the required
wavefunction or solver state, invokes the algorithm-specific iteration, and
performs any requested post-processing.
The new program driver can be exposed through a standalone executable
and, if needed, a facade in \texttt{python\_api.h} together with a
\texttt{pybind11} entry point. \texttt{OptOrbFCIProgram}
can be seen as an example that uses the same principle at a higher level: 
its dedicated program owns an outer orbital-optimization loop and invokes 
\texttt{CDFCIProgram} only as an inner solver.

\section{Results and Discussion}
\label{sec:experiments}

In this section, we conduct a series of numerical experiments to
evaluate the CDFCI software package on accuracy, robustness,
and scalability.

The first validation test is the Benzene Blind Test~\cite{eriksen_ground_2020},
where the goal is to determine the frozen-core ground-state energy of
the benzene molecule in a standard correlation-consistent basis set.
Several established selected CI methods (SHCI, ASCI, iCI) have been benchmarked
in the original paper; we compare our results with theirs.
Next, we evaluate a subset of the QUEST
database~\cite{Veril2021QUESTDB,Loos2018MountaineeringI,Loos2021QUESTDB},
which provides high-accuracy reference excitation energies for a broad range of
molecular systems.
After that, we turn to lattice Hamiltonians,
where we test a \(4\times 4\) Hubbard lattice with periodic boundary
conditions for different interaction strengths \(U/t=0.5,4,10\) and
different electron fillings \(\nelec = 14,15,16\). These results
are compared with reference data from exact diagonalization~\cite{Dagotto1992ED}.
Finally, we run computations for \ch{N2} in a basis set
comprising \(220\) spin--orbitals,
using different numbers of cores and demonstrate the
strong parallel scaling for large systems.

All the experiments were conducted on a system equipped with two
AMD EPYC 9754 128-core processors and 1.5 TB of memory.
The program is compiled using the GNU \texttt{g++} compiler version 11.4.0 with the
\texttt{-{}-O3} optimization flag and its native OpenMP support.
The orbitals and integrals are calculated via restricted
Hartree--Fock (RHF) in the PSI4 package~\cite{psi4} version 1.8. All energies are
reported in Hartree (Ha).

\subsection{Accuracy on Representative Benchmarks}

\subsubsection{Benzene Blind Test}

With the cc-pVDZ basis set,
the benzene molecule contains \(42\) electrons and \(114\) spatial orbitals.
After freezing the core electrons, the system still has \(30\) electrons and
\(108\) spatial orbitals, i.e., \(216\) spin--orbitals,
making it a challenging all-electron calculation.
In~\cite{eriksen_ground_2020},
the authors benchmarked a series of state-of-the-art FCI methods,
including SHCI, ASCI, iCI, DMRG, among others.
The results show that these methods provide qualitatively similar estimates of the correlation energy (around \(-863\) mHa),
which is defined as the difference between the total energy and the Hartree--Fock energy,
and usually used to quantify electron correlation effects.
Selected CI methods
employ a combination of variational estimation and second-order perturbative correction,
and extrapolate energies obtained under different thresholds to the FCI limit.
In this section, we perform calculations using the CDFCI software package on the same system with optimized orbitals~\footnote{Obtained from \url{https://github.com/seunghoonlee89/SI-benzene-paper-DMRG}.},
but we focus on the comparison of the variational energies and the number of determinants used.

\begin{table}[h]
  \caption{Results of the benzene benchmark test, including the correlation energy, number of Slater determinants included, and computational time, obtained using 64 cores and different thresholds \(\tau\).}
  \label{fig:benzene-conv}
  \centering
  \begin{tabular}{c c c c}
    \toprule
    \textbf{Threshold \(\tau\)} & \textbf{Correlation Energy (mHa)} & \textbf{Number of Determinants} & \textbf{Time (hours)} \\
    \midrule
    $0.01$ & $-716.18$ & $\num{181253}$ & $2.9$ \\
    $0.005$ & $-741.97$ & $\num{533932}$ & $5.8$ \\
    $0.004$ & $-752.84$ & $\num{893174}$ & $6.6$ \\
    $0.002$ & $-774.64$ & $\num{2870775}$ & $22.0$ \\
    $0.001$ & $-790.45$ & $\num{7513377}$ & $30.1$ \\
    $0.0005$ & $-806.12$ & $\num{20832092}$ & $37.3$ \\
    \bottomrule
  \end{tabular}
\end{table}

\begin{figure}[ht]
\centering
\begin{tikzpicture}
\begin{axis}[
  width=0.65\textwidth,
  height=0.45\textwidth,
  xlabel={Number of Slater Determinants},
  ylabel={Correlation Energy (mHa)},
  xmode=log,
  grid=both,
  tick label style={font=\small},
  label style={font=\small},
  legend style={
    at={(0.97,0.97)},
    anchor=north east,
    font=\small,
    draw=none
  },
]

\addplot+[
  mark=*,
  line width=1pt,
] coordinates {
(181253,   -716.18)
(533932,   -741.97)
(893174,   -752.84)
(2870775,  -774.64)
(7513377,  -790.45)
(20832092, -806.12)
};
\addlegendentry{CDFCI}

\addplot+[
  only marks,
  mark=square,
  mark size=3pt,
  color=orange,
] coordinates {
(1134081, -750.48)
(3901848, -773.17)
(13486304, -793.75)
(64100382, -815.64)
};
\addlegendentry{SHCI~\cite{eriksen_ground_2020}}

\addplot+[
  only marks,
  mark=star,
  mark size=4pt,
  color=green!60!black,
] coordinates {
(100000, -698.31)
(250000, -719.63)
(500000, -735.12)
(1000000, -749.43)
(2000000, -761.51)
(4000000, -772.35)
};
\addlegendentry{ASCI~\cite{eriksen_ground_2020}}

\end{axis}
\end{tikzpicture}
\caption{Relationship between the correlation energy and the number of Slater determinants in the variational space in the benzene benchmark test.
CDFCI results are compared with SHCI and ASCI reported data~\cite{eriksen_ground_2020}.}
\label{fig:benzene-E-vs-Ndet}
\end{figure}
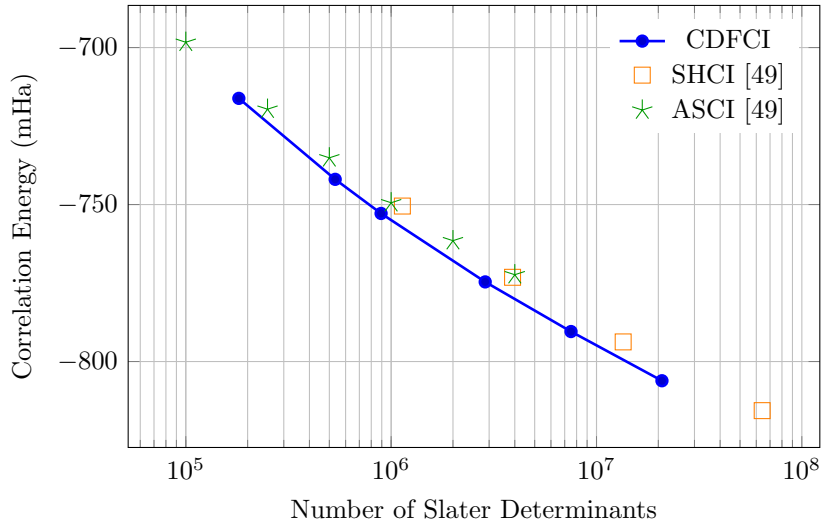

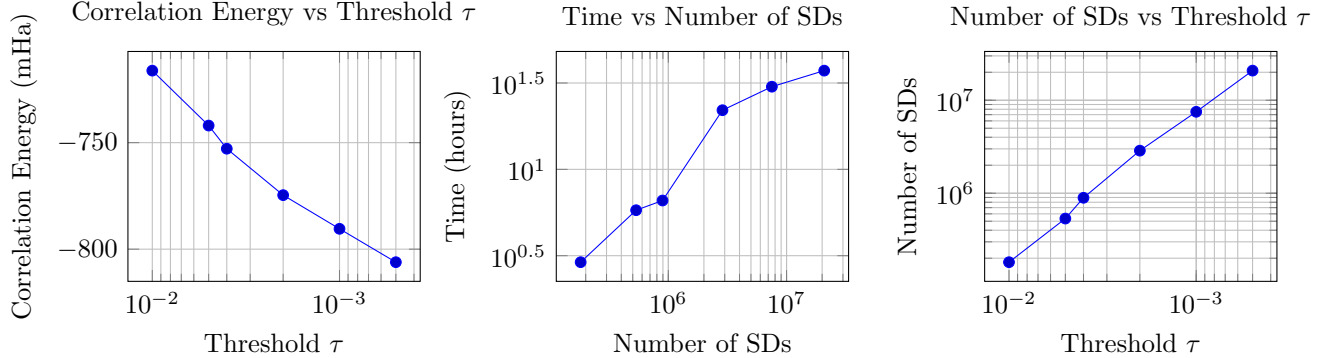
\begin{figure}[ht]
\centering
\begin{tikzpicture}
\begin{groupplot}[
  group style={group size=3 by 1, horizontal sep=1.5cm},
  width=0.31\textwidth,
  height=0.28\textwidth,
  xlabel style={font=\small},
  ylabel style={font=\small},
  tick label style={font=\small},
  title style={font=\small},
]

\pgfplotstableread{
tau      E        Ndet       time
0.01     -716.18  181253     2.9
0.005    -741.97  533932     5.8
0.004    -752.84  893174     6.6
0.002    -774.64  2870775    22.0
0.001    -790.45  7513377    30.1
0.0005   -806.12  20832092   37.3
}\datatable

\nextgroupplot[
  title={Correlation Energy vs Threshold $\tau$},
  xlabel={Threshold $\tau$},
  ylabel={Correlation Energy (mHa)},
  xmode=log,
  log basis x=10,
  grid=both,
  x dir=reverse,
]
\addplot+[
  mark=*,
] table[x=tau, y=E] {\datatable};

\nextgroupplot[
  title={Time vs Number of SDs},
  xlabel={Number of SDs},
  ylabel={Time (hours)},
  xmode=log,
  ymode=log,
  log basis x=10,
  log basis y=10,
  grid=both,
]
\addplot+[
  mark=*,
] table[x=Ndet, y=time] {\datatable};

\nextgroupplot[
  title={Number of SDs vs Threshold $\tau$},
  xlabel={Threshold $\tau$},
  ylabel={Number of SDs},
  xmode=log,
  ymode=log,
  log basis x=10,
  log basis y=10,
  x dir=reverse,
  grid=both,
]
\addplot+[
  mark=*,
] table[x=tau, y=Ndet] {\datatable};

\end{groupplot}
\end{tikzpicture}
\caption{Benzene benchmark test: The left plot shows the effect of threshold \(\tau\) on the correlation energy, the middle plot shows the effect of the number of Slater determinants (SDs) on computational time, and the right plot shows the effect of threshold \(\tau\) on the number of SDs.}
\label{fig:benzene-plots}
\end{figure}

Table~\ref{fig:benzene-conv} shows
the variational correlation energy obtained by CDFCI under different thresholds \(\tau\),
together with the number of determinants used and the computational time.
Clearly, as \(\tau\) decreases,
the correlation energy gradually decreases,
while the number of determinants and computational time increase significantly.
For the most time-consuming calculation (\(\tau = 0.0005\)),
CDFCI achieves a correlation energy of \(-806.12\) mHa
using approximately \(2.08 \times 10^7\) determinants in about \(37.3\) hours,
which corresponds to approximately \(2.4 k\) core-hours,
and is comparable to other CI methods reported in~\cite{eriksen_ground_2020}.

Figure~\ref{fig:benzene-E-vs-Ndet} shows an approximately linear relationship
between the correlation energy and the number of determinants for CDFCI under different thresholds \(\tau\) (blue solid line).
For comparison, we also include reference data for SHCI and ASCI from~\cite{eriksen_ground_2020}.
Both SHCI data points (orange squares) and ASCI data points (green stars) indicate that, for a similar number of determinants,
the correlation energy obtained by SHCI or ASCI is slightly higher than that of CDFCI.
This observation demonstrates the efficiency of CDFCI in capturing important determinants,
allowing it to achieve better correlation energy estimates with fewer determinants,
thus validating the effectiveness of gradient- or residual-based determinant selection strategies.

Figure~\ref{fig:benzene-plots} further analyzes different aspects of the CDFCI results.
The left plot shows the decrease of correlation energy with respect to \(\tau\),
and exhibits an approximately linear relationship.
The middle plot shows the relationship between computational time and the number of determinants,
indicating that the growth of computational time approximately follows a power-law trend.
The right plot shows the relationship between the number of Slater determinants and \(\tau\),
where the number of determinants increases approximately following a power-law trend as \(\tau\) decreases,
consistent with the time growth observed in the middle plot.

\subsubsection{QUEST Database Subset}
To evaluate the performance of xCDFCI in excited-state calculations,
we selected three representative molecules from the QUEST dataset~\cite{Loos2018MountaineeringI},
covering different types of excitations,
including singlet states, triplet states,
valence states, and Rydberg states.
This benchmark aims to demonstrate that xCDFCI can achieve systematic and reliable accuracy
across different excitation types.
We report the final vertical excitation energies
for each state,
as well as the exFCI reference results, obtained via extrapolation from FCI results.
The geometries used in the QUEST dataset are optimized at the CC3/aug-cc-pVTZ level.
We use the same geometries to perform Hartree--Fock calculations to obtain one- and two-body integrals
that are subsequently used in the xCDFCI calculations.

\begin{table}[h]
  \caption{Vertical excitation energies of excited states for \ch{H2O}/aug-cc-pVTZ in the QUEST dataset,
  compared with exFCI reference data~\cite{Loos2018MountaineeringI},
  along with the number of Slater determinants (SDs) used in exFCI.
  In the xCDFCI calculations, the number of determinants used is \(\num{4935670}\).}
  \label{tab:quest-h2o}
  \centering
  \begin{tabular}{cccc}
    \toprule
    \textbf{Excited State} & \makecell{\textbf{Vertical Excitation}\\\textbf{Energy (eV)}} & \makecell{\textbf{exFCI Reference}\\\textbf{(eV)}} & \makecell{\textbf{Number of SDs}\\\textbf{in exFCI}} \\
    \midrule
    $^3B_1 (n \to 3s)$ & $7.26$ & $7.24$ & $\num{5950423}$ \\
    $^1B_1 (n \to 3s)$ & $7.63$ & $7.62$ & $\num{5589200}$ \\
    $^3A_2 (n \to 3p)$ & $9.29$ & $9.24$ & $\num{3760373}$ \\
    $^1A_2 (n \to 3p)$ & $9.43$ & $9.41$ & $\num{5589200}$ \\
    $^3A_1 (n \to 3s)$ & $9.55$ & $9.54$ & $\num{3760373}$ \\
    $^1A_1 (n \to 3s)$ & $9.99$ & $9.99$ & $\num{5589200}$ \\
    \bottomrule
  \end{tabular}
\end{table}

Water (\ch{H2O}), due to its simple structure and important chemical properties,
is a classical test system for Rydberg excitation calculations.
We compute the ground state and six excited states of water using the aug-cc-pVTZ basis set,
including three triplet states and three singlet states,
involving excitations from the nonbonding \(n\) orbital to Rydberg \(3s\) and \(3p\) orbitals.
To compare with reference data,
we use a threshold of \(\tau=0.001\)
to control the computational space within the same order of magnitude
(approximately \(\num{4935670}\) determinants).
The xCDFCI program runs in parallel using 8 threads and converges in about one day.
Table~\ref{tab:quest-h2o} summarizes
the final vertical excitation energies and their errors relative to the reference data.
It can be seen that the excitation energies obtained by xCDFCI are very close to the exFCI reference values.
Except for the $^3A_2 (n \to 3p)$ state,
the errors of the other five excited states are within \(0.02\) eV.
Moreover, the excitation energies obtained by xCDFCI are generally lower than the exFCI reference values.
Since Rydberg states are more sensitive to the basis set,
the correlation effects captured in this system may lead to a systematic lowering of excitation energies,
bringing them closer to results obtained with the aug-cc-pVQZ basis set.

\begin{table}[h]
  \caption{Vertical excitation energies of excited states for \ch{N2}/aug-cc-pVDZ in the QUEST dataset,
  compared with exFCI reference data~\cite{Loos2018MountaineeringI},
  along with the number of Slater determinants (SDs) used in exFCI.
  In the xCDFCI calculations, the number of determinants used is \(\num{20558436}\).}
  \label{tab:quest-n2}
  \centering
  \begin{tabular}{cccc}
    \toprule
    \textbf{Excited State} & \makecell{\textbf{Vertical Excitation}\\\textbf{Energy (eV)}} & \makecell{\textbf{exFCI Reference}\\\textbf{(eV)}} & \makecell{\textbf{Number of SDs}\\\textbf{in exFCI}} \\
    \midrule
    $^3\Sigma_u^+ (\pi \to \pi^*)$ & $7.67$ & $7.70$ & $\num{8139401}$ \\
    $^3\Pi_g (n \to \pi^*)$        & $8.08$ & $8.05$ & $\num{2705349}$ \\
    $^3\Delta_u (\pi \to \pi^*)$   & $8.94$ & $8.96$ & $\num{2705349}$ \\
    $^1\Pi_g (n \to \pi^*)$        & $9.44$ & $9.41$ & $\num{2775773}$ \\
    $^3\Sigma_u^- (\pi \to \pi^*)$ & $9.73$ & $9.75$ & $\num{2705349}$ \\
    $^1\Sigma_u^- (\pi \to \pi^*)$ & $10.03$ & $10.05$ & $\num{2775773}$ \\
    $^1\Delta_u (\pi \to \pi^*)$   & $10.41$ & $10.43$ & $\num{2775773}$ \\
    \bottomrule
  \end{tabular}
\end{table}

Nitrogen (\ch{N2}) is a strongly correlated system.
We compute its ground state and seven excited states using the aug-cc-pVDZ basis set.
Among them, the $^3\Delta_u (\pi \to \pi^*)$ and
$^1\Delta_u (\pi \to \pi^*)$ states are doubly degenerate,
so in the actual xCDFCI calculations,
we compute nine excited states to cover all degeneracies.
We use a threshold of \(\tau=0.004\)
to control the computational space within a comparable size
(approximately \(\num{20558436}\) determinants).
The xCDFCI program runs in parallel using 8 threads and converges in about 12 hours.
Table~\ref{tab:quest-n2} summarizes
the final vertical excitation energies and their errors relative to the reference data.
It can be seen that the deviations of xCDFCI excitation energies from exFCI reference values
are within \(0.03\) eV.
Unlike the water system,
there is no clear systematic bias in the energy differences for this system.

\begin{table}[h]
  \caption{Vertical excitation energies of excited states for \ch{C2H2}/aug-cc-pVDZ in the QUEST dataset,
  compared with exFCI reference data~\cite{Loos2018MountaineeringI},
  along with the number of Slater determinants (SDs) used in exFCI.
  In the xCDFCI calculations, the number of determinants used is \(\num{27874551}\).}
  \label{tab:quest-c2h2}
  \centering
  \begin{tabular}{cccc}
    \toprule
    \textbf{Excited State} & \makecell{\textbf{Vertical Excitation}\\\textbf{Energy (eV)}} & \makecell{\textbf{exFCI Reference}\\\textbf{(eV)}} & \makecell{\textbf{Number of SDs}\\\textbf{in exFCI}} \\
    \midrule
    $^3\Sigma_u^+ (\pi \to \pi^*)$ & $5.51$ & $5.50$ & $\num{8494075}$ \\
    $^3\Delta_u (\pi \to \pi^*)$   & $6.46$ & $6.46$ & $\num{4403434}$ \\
    $^3\Sigma_u^- (\pi \to \pi^*)$ & $7.13$ & $7.14$ & $\num{4403434}$ \\
    $^1\Sigma_u^- (\pi \to \pi^*)$ & $7.20$ & $7.20$ & $\num{4162848}$ \\
    $^1\Delta_u (\pi \to \pi^*)$   & $7.51$ & $7.51$ & $\num{4162848}$ \\
    \bottomrule
  \end{tabular}
\end{table}

Acetylene (\ch{C2H2})
is the smallest conjugated organic molecule with stable low-lying excited states,
and is therefore important for studying vertical fluorescence transitions.
We compute its five lowest-energy valence excited states in a linear geometry,
including two doubly degenerate \(\Delta\) states.
We use a threshold of \(\tau=0.0002\)
to control the computational space within the same order of magnitude
(approximately \(\num{27874551}\) determinants).
The xCDFCI program runs in parallel using 8 threads and converges in about $36$ hours.
Table~\ref{tab:quest-c2h2} summarizes
the final vertical excitation energies and their errors relative to the reference data.
It can be seen that, although it uses more determinants than exFCI,
the excitation energies obtained by xCDFCI agree very well with exFCI reference values,
with deviations within \(0.01\) eV.

\subsubsection{2D Hubbard Model}
We consider a \(4\times 4\) Hubbard lattice with periodic boundary conditions (PBC),
with interaction strength \(U/t=4\),
and electron fillings \(\nelec = 14,15,16\).
These results demonstrate the capability of CDFCI in handling lattice Hamiltonians,
and are compared with reference data from exact diagonalization~\cite{Dagotto1992ED}.

The initialization for the Hubbard model is more complicated than that for molecular systems.
In molecular calculations, one typically starts from the Hartree--Fock reference state,
in which electrons occupy the orbitals with lowest energies.
The Hartree--Fock
procedure guarantees the presence of an energy gap between the highest occupied molecular orbital (HOMO)
and the lowest unoccupied molecular orbital (LUMO).
In contrast, for the Hubbard model, localized site states can be degenerate in energy levels,
leading to multiple configurations that share the same minimal energy.
In the current implementation, we identify all such degenerate lowest-energy configurations
and extract the corresponding submatrix of the Hamiltonian.
The eigenvector associated with the smallest eigenvalue of this submatrix is then used as the initial state for the CDFCI algorithm.
For example, in the \(4 \times 4\) PBC Hubbard model,
\(16\) site states consist of \(5\) states with negative energy,
\(6\) with zero energy, and \(5\) with positive energy.
At half filling (\(\nelec = 16\)) and under the constraint of total spin zero,
there are \(\binom{6}{3}^2 = 20^2 = 400\) distinct configurations that all attain the same lowest energy.
We construct the Hamiltonian submatrix corresponding to these \(400\) determinants
and compute its ground-state eigenvector as the initialization.
Numerical experiments show that this strategy yields significantly better performance
than alternative initialization approaches for this system.

\begin{figure}[htbp]
\centering
\begin{tikzpicture}
\begin{axis}[
  width=0.65\textwidth,
  height=0.45\textwidth,
    xmode=linear,
    ymode=log,
    ymin=1.1e-4,
    ymax=5e-1,
    xlabel={Iterations},
    ylabel={Energy Difference (log scale)},
    title={Hubbard Model Ground State Energy Convergence ($U=4$, $4\times4$ Lattice)},
    grid=both,
    legend pos=north east,
    line width=1pt,
    tick align=outside,
    scaled ticks=false,
]
\addplot[
    color=blue,
]
table[x=iteration,y=energy_diff,col sep=space] {data/hubbard_n16.dat};
\addlegendentry{n=16}

\addplot[
    color=orange,
]
table[x=iteration,y=energy_diff,col sep=space] {data/hubbard_n15.dat};
\addlegendentry{n=15}

\addplot[
    color=green!60!black,
]
table[x=iteration,y=energy_diff,col sep=space] {data/hubbard_n14.dat};
\addlegendentry{n=14}
\end{axis}
\end{tikzpicture}
  \caption{Ground state energy convergence curves of the two-dimensional Hubbard model with periodic boundary conditions
  at $U/t=4$ for different electron fillings.
  Reference data are from exact diagonalization~\cite{Dagotto1992ED}.}
  \label{fig:hubbard}
\end{figure}
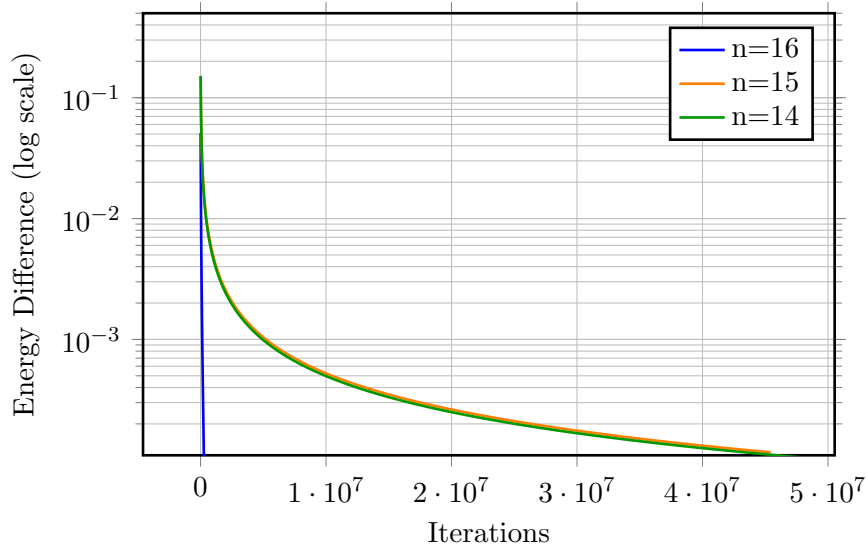

Figure~\ref{fig:hubbard} shows the ground-state energy convergence curves
of the two-dimensional Hubbard model with periodic boundary conditions
at \(U/t=4\) for different electron fillings \(\nelec = 14,15,16\).
For \(\nelec = 14\),
CDFCI converges to an energy of \(-15.74449\) Ha
within a variational space of approximately \(8\times 10^6\) determinants,
with a deviation of about \(0.1\) mHa from the exact diagonalization result \(-15.74459\) Ha.
For \(\nelec = 15\),
CDFCI converges to an energy of \(-14.66512\) Ha
within a variational space of approximately \(9\times 10^6\) determinants,
with a deviation of about \(0.1\) mHa from the exact diagonalization result \(-14.66524\) Ha.
For \(\nelec = 16\),
CDFCI converges to an energy of \(-13.62185\) Ha
with approximately \(1\times 10^7\) determinants,
which is identical to the exact diagonalization result.
From the figure, it can be observed that
for \(\nelec = 14\) and \(\nelec = 15\),
the convergence of the energy is relatively slow,
whereas for \(\nelec = 16\), the convergence is significantly faster.
The convergence behavior of the ground-state energy in the Hubbard model
is closely related to the electron filling.
For \(\nelec = 14\) and \(\nelec = 15\),
the system is in a lightly doped regime with strong electron correlation,
leading to slower convergence.
For \(\nelec = 16\),
the system is at half filling with weaker correlation effects,
resulting in faster convergence.

\begin{table}
  \caption{Ground state energies and computational time for the two-dimensional $4\times 4$ half-filled Hubbard model with periodic boundary conditions at \(U/t=0.5, 4, 10\). Reference data are from exact diagonalization~\cite{Dagotto1992ED}.}
  \label{tab:hubbard-ut}
  \centering
  \begin{tabular}{cccc}
    \toprule
    \textbf{$U/t$} & \textbf{CDFCI Energy (Ha)} & \textbf{Exact Diagonalization Energy (Ha)} & \textbf{Time} \\
    \midrule
    $0.5$ & $-22.34023$ & $-22.34023$ & $13$ seconds \\
    $4$ & $-13.62185$ & $-13.62185$ & $9$ minutes \\
    $10$ & $-7.02682$ & $-7.02900$ & $12$ hours \\
    \bottomrule
  \end{tabular}
\end{table}

Table~\ref{tab:hubbard-ut} summarizes the comparison of ground-state energies and computational time
for the two-dimensional \(4\times 4\) Hubbard model with periodic boundary conditions
at half filling (\(\nelec = 16\)) under different \(U/t\) values.
Significant differences in computational time can be observed.
For weakly correlated systems (\(U/t=0.5\)),
CDFCI converges to the exact diagonalization energy within about \(13\) seconds.
For moderately correlated systems (\(U/t=4\)),
CDFCI converges within about \(9\) minutes.
For strongly correlated systems (\(U/t=10\)),
CDFCI converges to \(-7.02682\) Ha within about \(12\) hours,
with a deviation of approximately \(2.2\) mHa from the exact diagonalization result \(-7.02900\) Ha.
These results indicate that CDFCI performs well across different correlation regimes of the Hubbard model.
However, for strongly correlated systems, energy convergence requires longer computational time,
suggesting that more efficient optimization algorithms may be needed
to accelerate convergence in strongly correlated regimes.

\subsection{Scalability}
Finally, we evaluate the parallel scalability of CDFCI under different computational resources.
We fix the problem size and workload
to evaluate the strong scaling performance of CDFCI as computational resources increase.
For the \ch{N2}/cc-pVQZ system, we
run CDFCI ground state calculations on different numbers of CPU cores,
and record the total runtime and speedup for each calculation.
We test \(8\), \(16\), \(32\), \(64\),
\(128\), and \(256\) CPU cores.
The threshold is set to \(\tau=5\times 10^{-5}\),
with \(128\) coordinates updated in each iteration,
and after \(\num{400000}\) iterations the ground-state energy obtained is
\(-109.46134\) Ha.
This result differs from the fully converged energy \(-109.46256\) Ha
by about \(1.3\) mHa, which is within chemical accuracy.
We set the wavefunction memory budget to \(500\) GiB, which results in a
preallocated hash table capacity of \(2^{33}\) slots. The peak resident set
size measured in the 32-thread run was approximately \(403\) GiB, consistent
with the memory estimate in Section~\ref{sec:computational-complexity} based
on the allocated hash table capacity.

\begin{table}[h]
  \caption{Runtime and speedup of CDFCI on the \ch{N2}/cc-pVQZ system using different numbers of CPU cores.
  The speedup is defined as $8 \times T_8 / T_N$, where $T_8$ is the runtime using 8 cores and $T_N$ is the runtime using $N$ cores.}
  \label{tab:scalability}
  \centering
  \begin{tabular}{c c c}
    \toprule
    \textbf{Number of CPU Cores} & \textbf{Runtime (seconds)} & \textbf{Speedup} \\
    \midrule
    $8$   & $116904.5$ & $8.0$ \\
    $16$  & $54128.3$ & $17.3$ \\
    $32$  & $27514.0$ & $34.0$ \\
    $64$  & $15913.6$  & $58.8$ \\
    $128$ & $12566.7$  & $74.4$ \\
    $256$ & $9375.6$  & $99.8$ \\
    \bottomrule
  \end{tabular}
\end{table}

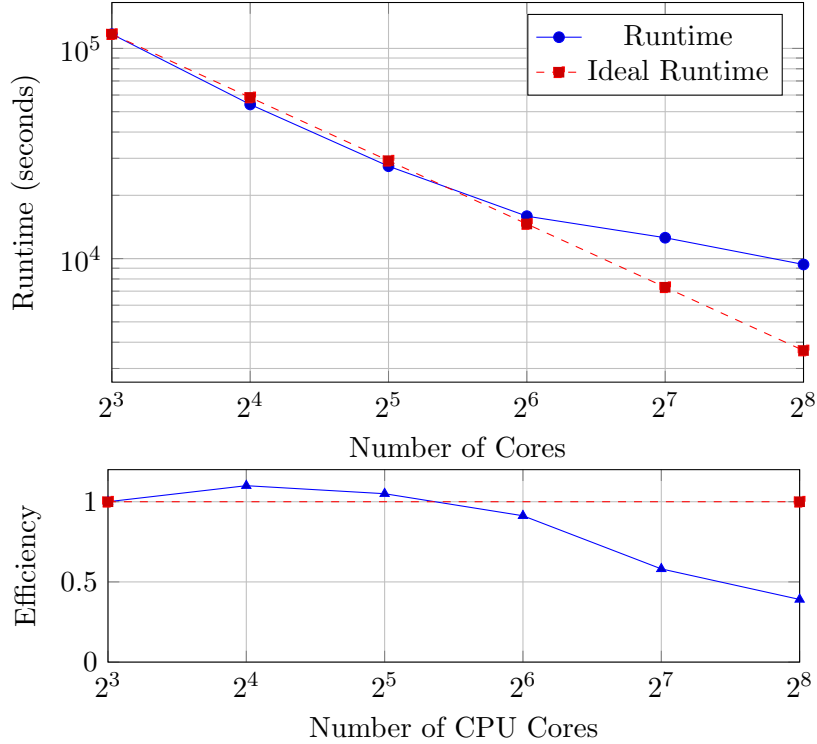
\begin{figure}[h]
\centering
\begin{tikzpicture}

\begin{axis}[
  width=0.65\textwidth,
  height=0.4\textwidth,
    xmode=log,
    log basis x=2,
    xtick={8,16,32,64,128,256},
    xmin=8, xmax=256,
    xlabel={Number of Cores},
    ylabel={Runtime (seconds)},
    ymode=log,
    log basis y=10,
    legend style={at={(0.97,0.97)},anchor=north east},
    grid=both,
    anchor=north east,
]

\addplot+[mark=*] coordinates {
(8,116904.5)
(16,54128.3)
(32,27514.0)
(64,15913.6)
(128,12566.7)
(256,9375.6)
};
\addlegendentry{Runtime}

\addplot+[dashed] coordinates {
(8,116904.5)
(16,58452.25)
(32,29226.125)
(64,14613.0625)
(128,7306.53125)
(256,3653.265625)
};
\addlegendentry{Ideal Runtime}

\end{axis}

\end{tikzpicture}

\begin{tikzpicture}
\begin{axis}[
  width=0.65\textwidth,
  height=0.25\textwidth,
    xmode=log,
    log basis x=2,
    xtick={8,16,32,64,128,256},
    xmin=8, xmax=256,
    xlabel={Number of CPU Cores},
    ylabel={Efficiency},
    ymin=0, ymax=1.2,
    grid=both,
]

\addplot+[mark=triangle*] coordinates {
(8,1.0)
(16,1.1)
(32,1.05)
(64,0.9125)
(128,0.58125)
(256,0.390625)
};

\addplot+[dashed] coordinates {(8,1) (256,1)};

\end{axis}
\end{tikzpicture}

\caption{Strong scaling of CDFCI on the \ch{N2}/cc-pVQZ system:
the top panel shows runtime (log scale) and ideal linear scaling,
and the bottom panel shows efficiency, defined as speedup divided by the number of processors.}
\label{fig:scalability_time_speedup_eff}
\end{figure}

Table~\ref{tab:scalability} summarizes the runtime and speedup
for different numbers of CPU cores.
It can be seen that, as the number of CPU cores increases, the runtime decreases significantly,
and even superlinear speedup is observed at \(16\) and \(32\) cores.
However, when the number of CPU cores increases beyond \(64\),
the growth in speedup begins to diminish noticeably.

Figure~\ref{fig:scalability_time_speedup_eff} provides a more intuitive illustration.
The top panel shows the decrease in runtime as the number of CPU cores increases, compared with the ideal linear scaling.
The bottom panel presents the parallel efficiency, defined as the speedup divided by the number of processors.
When the number of CPU cores increases from 8 to 16 and 32,
a slight superlinear speedup can be observed.
This phenomenon is usually attributed to improved cache utilization
and memory hierarchy effects.
More specifically, as the workload per thread decreases,
the effective working set becomes smaller
and fits better into the last-level cache,
thereby reducing memory latency and improving effective bandwidth.
Since the CDFCI algorithm is mainly memory-bound due to irregular hash table accesses
and Hamiltonian evaluation,
improved cache locality can lead to noticeable superlinear scaling
at moderate CPU core counts.
However, when the number of CPU cores exceeds 64,
the speedup gradually saturates.
This indicates that memory bandwidth becomes a limiting factor
and synchronization overhead increases,
which is common in large-scale shared-memory parallelization of sparse algorithms.
Overall, CDFCI demonstrates good parallel scalability on the \ch{N2}/cc-pVQZ system,
especially for moderate numbers of CPU cores,
although the performance gain becomes limited
in larger-scale parallel computations.

\section{Conclusion and Future Work}
\label{sec:conclusion}
We have presented CDFCI, an efficient and modular software package for
solving the non-relativistic, time-independent
fermionic Schr\"odinger equation
within the full configuration interaction framework.
CDFCI leverages coordinate descent methods to iteratively refine
the wavefunction representation, achieving high accuracy with
reduced computational cost.
With appropriately defined objective functions,
the framework can compute multiple states simultaneously.
The software architecture is modular and extensible,
facilitating integration of new Hamiltonian types, wavefunction storage
backends, and solver algorithms.
Shared-memory parallelization enables efficient utilization of modern multi-core processors.
Benchmark results demonstrate millihartree-level agreement with reference 
excitation energies for the three QUEST molecules and with reference 
ground-state energies for the tested Hubbard models, with sub-millihartree 
agreement achieved in multiple cases.

Future work will focus on incorporating parallel linear
algebra libraries to enhance performance,
and exploring more advanced perturbation theory
techniques to further improve
the accuracy of the computed energies.
Another direction is to go beyond non-relativistic fermionic Hamiltonians,
that is, to extend the framework to bosonic systems and
relativistic Hamiltonians. This will broaden the applicability
of CDFCI and position it as a general solver across
quantum chemistry and condensed matter physics.

\section*{Acknowledgments}
This work was supported in part
by the National Natural Science Foundation of China under Grant
Nos.~12271109 and 12526211;
by the U.S. National Science Foundation
(NSF) under Grant Nos.~DMS-1454939 and DMS-2012286;
by the Shanghai Pilot Program for Basic Research-Fudan University
under Grant No.~21TQ1400100 (22TQ017);
by the Scientific Research Innovation Capability Support Project
for Young Faculty under Grant No.~SRICSPYF-ZY2025159;
and by the Xuemin Institute of Advanced Studies, Fudan University.

\bibliographystyle{unsrt}
\bibliography{main}

\appendix

\section[Determining Step Size]{Determining Step Size in Problem~\eqref{eq:choose-alpha}
and~\eqref{eq:choose-eta}}
\label{app:line-search}
We describe in detail how to minimize the quartic polynomial
in Problem~\eqref{eq:choose-alpha} and~\eqref{eq:choose-eta},
thereby determining the optimal step size \(\eta\).

In Problem~\eqref{eq:choose-alpha},
we have
\begin{equation}
  h(\eta) = f(\currvc + \eta \vve_{\ichosen}) = \eta^4 + v^{(\ell)}_3 \eta^3
   + v^{(\ell)}_2 \eta^2 + v^{(\ell)}_1 \eta + v^{(\ell)}_0,
\end{equation}
with
\begin{equation}
  \begin{split}
    v^{(\ell)}_0 &= f(\currvc), \\
    v^{(\ell)}_1 &= 4 \currvb_{\ichosen} + 4 \ccnorm^{(\ell)}
    \currvc_{\ichosen}, \\
    v^{(\ell)}_2 &= 2 H_{\ichosen, \ichosen} + 4 (\currvc_{\ichosen})^2
   + 2 \ccnorm^{(\ell)}, \\
    v^{(\ell)}_3 &= 4 \currvc_{\ichosen}.
  \end{split}
\end{equation}
The stationary condition
\(\frac{\diff h(\eta)}{\diff \eta} = 0\) leads to a cubic equation
of the form
\begin{equation}
  \label{eq:cubic-v}
  \frac{\diff h(\eta)}{\diff \eta} = 4\eta^3 + 3v^{(\ell)}_3 \eta^2
  + 2v^{(\ell)}_2 \eta + v^{(\ell)}_1 = 0.
\end{equation}

In Problem~\eqref{eq:choose-eta},
we have
\begin{equation}
  h(\eta) = f(\currC + \eta
  \vve_{\ichosen}\currG_{\ichosen, :} )
   = w^{(\ell)}_4\eta^4 + w^{(\ell)}_3 \eta^3
   + w^{(\ell)}_2 \eta^2 + w^{(\ell)}_1 \eta + w^{(\ell)}_0,
\end{equation}
with
\begin{equation}
  \begin{split}
    w^{(\ell)}_0 &= f(\currC), \\
    w^{(\ell)}_1 &= 4 \left\|\currG_{\ichosen, :}\right\|^2, \\
    w^{(\ell)}_2 &= 2 H_{\ichosen, \ichosen} \left\|\currG_{\ichosen, :}\right\|^2
     + 2 \left(\currC_{\ichosen, :}
     \left(\currG\right)^\T_{\ichosen, :}\right)^2
     + 2 \left\|\currC_{\ichosen, :}\right\|^2 \left\|\currG_{\ichosen, :}\right\|^2 \\
     & \quad + 2 \currG_{\ichosen, :} \CCnorm^{(\ell)}
     \left(\currG\right)^\T_{\ichosen, :}, \\
    w^{(\ell)}_3 &= 4 \left\|\currG_{\ichosen, :}\right\|^2 \currC_{\ichosen,:}
    \left(\currG\right)^\T_{\ichosen, :}, \\
    w^{(\ell)}_4 &= \left\|\currG_{\ichosen, :}\right\|^4.
  \end{split}
\end{equation}
The stationary condition
\(\frac{\diff h(\eta)}{\diff \eta} = 0\) leads to a cubic equation
of the form
\begin{equation}
  \label{eq:cubic-w}
  \frac{\diff h(\eta)}{\diff \eta} = 4w^{(\ell)}_4\eta^3 + 3w^{(\ell)}_3 \eta^2
  + 2w^{(\ell)}_2 \eta + w^{(\ell)}_1 = 0.
\end{equation}

To solve the cubic equations~\eqref{eq:cubic-v} and~\eqref{eq:cubic-w},
we use Cardano's method~\cite{nickalls1993cardano}. The method first converts the cubic
equation into a depressed cubic form \(x^3 + px + q = 0\) via the substitution
\(\eta = x - \frac{b}{3a}\),
where \(a\), \(b\), \(c\), and \(d\) are the coefficients of the cubic equation.
The coefficients \(p\) and \(q\) are computed as
\begin{equation}
  p = \frac{3ac - b^2}{3a^2}, \quad
  q = \frac{2b^3 - 9abc + 27a^2d}{27a^3}.
\end{equation}
The discriminant \(\Delta = \left(\frac{q}{2}\right)^2 + \left(\frac{p}{3}\right)^3\)
determines the number of real roots.
When $\Delta \ge 0$,
the single real root without multiplicity is selected directly.
When $\Delta < 0$, the cubic equation has three distinct real roots
$x_1 < x_2 < x_3$, and the function exhibits one local maximum
and one local minimum between them.
The root further away from the middle one minimizes the quartic polynomial.

After obtaining an analytical root, we refine it using Newton's iteration:
\[
x_{k+1} = x_k - \frac{f(x_k)}{f'(x_k)},
\]
until the relative change satisfies
$|x_{k+1} - x_k| / |x_k| < \varepsilon$.
Finally, set \(\eta = x_{k+1}\).

\end{document}